\documentclass[preprint,12pt]{elsarticle}

\usepackage{amsmath,amssymb,amsfonts,mathrsfs,bm}

\usepackage{amsmath,amssymb,amsfonts,mathrsfs,bm}
\usepackage{indentfirst}
\usepackage{graphicx}
\usepackage{float,subfig}

\newtheorem{remark}{Remark}[section]

\newcommand{\ignore}[1]{}

\makeatletter
\@addtoreset{equation}{section}
\makeatother

\makeatletter
\def\ps@pprintTitle{%
	\let\@oddhead\@empty
	\let\@evenhead\@empty
	\def\@oddfoot{\reset@font\hfil\thepage\hfil}
	\let\@evenfoot\@oddfoot
}
\makeatother

\allowdisplaybreaks
\date{}
\begin{document}

\begin{frontmatter}

\title{Fixed-time synchronization for quaternion-valued memristor-based neural networks with mixed delays
}

\author[label1]{Yanlin Zhang}
 \ead{yanlnzhang@163.com}
 \author[label1]{Liqiao Yang}
 \ead{liqiaoyoung@163.com}
 \author[label1]{Kit Ian Kou\corref{cor1}}
 \ead{kikou@umac.mo}
 \author[label2]{ Yang Liu}
  \ead{liuyang@zjnu.edu.cn}

\affiliation[label1]{organization={Department of Mathematics, Faculty of Science and Technology},
            addressline={University of Macau}, 
            city={Macau},
            postcode={999078}, 
            country={China}}
        
\affiliation[label2]{organization={ College of Mathematics and Computer Science},
        	addressline={Zhejiang Normal University},
        	city={Jinhua},
        	postcode={321004},
        	country={China}}
        
\cortext[cor1]{Corresponding author}

\begin{abstract}
In this paper, the fixed-time synchronization (FXTSYN) of unilateral coefficients quaternion-valued memristor-based neural networks (UCQVMNNs) with mixed delays is investigated. Instead of decomposition, a direct analytical method is proposed to achieve FXTSYN of UCQVMNNs using one-norm smoothly. 
Then apply the set-valued map and the differential inclusion theorem to handle discontinuity problems of drive-response systems. The novel nonlinear controllers together with the Lyapunov function are designed to achieve the control goal. Using the FXTSYN theory and inequality techniques, some criteria of FXTSYN for UCQVMNNs are given. Furthermore, the estimated settling time is obtained explicitly. Finally, numerical simulations are presented to demonstrate the correctness, effectiveness and practicability of the obtained theoretical results.

\end{abstract}


\end{frontmatter}


\section{Introduction}
\label{}
Quaternion is a subset of Clifford algebra, which was invented by Hamilton in 1843 \cite{hamilton1848xi} and is a natural extension of complex space. The state variables, input variables, connection weights, and the activation functions of quaternion-valued neural networks (QVNNs) all take values in the quaternion field. Moreover, QVNNs have been used in a variety of practical applications, such as satellite TV, aerospace, 3-D wind processing, color image processing, polarized waves, and space rotation \cite{ali2021global, took2008quaternion, yang2022quaternion, zou2016quaternion}.
However, in contrast to real- and complex-values, some arithmetic rules such as the commutativity of multiplication do not apply to quaternion. As a result, these researches on NNs are primarily focused on the real-valued and complex-valued fields \cite{ding2017robust, jankowski1996complex, zhang2019finite, zhou2017finite}, while the related research on quaternion is relatively scarce in recent decades.

Due to the non-commutative of quaternion multiplication, there are a great variety of polynomials in quaternion algebra than in the real and complex fields, such as polynomials with left, right, and bilateral coefficients. In QDEs, those equations with bilateral coefficients are too difficult to solve, so there are not many results \cite{wilczynski2009quaternionic, cai2018solving} about it. Therefore, in this paper, we first study the unilateral coefficients of quaternion-valued neural networks (UCQVNNs).

As one of the basic circuit elements, the memristor (an abbreviation for the memory and the resistor) was first proposed by Chua \cite{chua1971memristor} in 1971.
The memristor has the properties of memory and nanoscale, which can better and more realistically simulate the biological synapse. It describes the relationship between the charge and the magnetic flux, so the memristor systems are more precise models of artificial neural networks. Furthermore, it has the potential to improve the application of pattern recognition, combinatorial optimization, and data processing \cite{cantley2011hebbian, corinto2011nonlinear, itoh2019memristor}. In recent years, many scholars have studied many characteristics of memristor-based neural networks (MNNs) due to the powerful functions in human brain computers \cite{DBLP:journals/nn/AbdurahmanJT15, chen2017fixed, fu2018dissipativity, li2015lag, wei2018finite}, such as dissipativity, stability, synchronization and so on.

It is widely known that synchronization is an expansion of stability, which is the dynamic behavior of the drive-response system to achieve the same state at the same time. Due to the powerful role of synchronization in non-linear systems \cite{balasubramaniam2011synchronization}, it is widely used in areas such as information processing \cite{singer1993synchronization}, security communication \cite{yang2014exponential}, chemical reaction \cite{parlitz1992transmission} and so on.
Our research aims to achieve synchronization as quickly as possible under the controller's action. Fortunately, the fixed time synchronization (FXTSYN) \cite{polyakov2011nonlinear} was quickly proposed, in which the settling time is completely irrelevant to initial values. And FXTSYN has a fast convergence time and better interference inhibitory characteristics. Indeed, many practical phenomena, such as information processing and biological systems, require rapid synchronization to maintain normal order. Therefore, many scholars pay attention to the FXTSYN of real-valued MNNs (RVMNNs) and complex-valued MNNs (CVMNNs), some promising applications can be found in
\cite{wei2018finite, DBLP:journals/chinaf/CaoL17, chen2018fixed, feng2021fixed, guo2017finite, kashkynbayev2022finite}.

However, as far as we know, compared to the fields of RVMNNs and CVMNNs, there are quite rare results on FXTSNY of UCQVMNNs with mixed delays \cite{chen2020fixed, deng2019fixed, li2020finite, peng2022finite, wei2019fixed}. And the method used by most of them is the decomposition method, in \cite{chen2020fixed}, the FXTSNY of a class of QVNNs with memristor was investigated via decomposition methods. That is, the real and imaginary parts of a quaternion neural network are re-expressed as four real-valued systems, which lose important information about the original problem structure. Regrettably, it is generally accompanied by a more complex derivation process, and it may also increase conservatism. Therefore, finding an easier and no-decomposition method to research the FXTSNY of MNNs is necessary and important. In \cite{peng2022finite}, Peng et al. use a direct analytical method to study the FNTSNY and FXTSNY problems of QVNNs by introducing an improved one-norm, which is without decomposition techniques. Compared with the traditional decomposing method, even more specific FXTSNY is achieved by the one-norm method, which shows the strong applicability and less conservative of this method. So, we design two novel nonlinear controllers based on the improved one-norm method to achieve the FXTSNY of UCQVMNNs with mixed delays in this paper. It is of great significance and value for research.

Motivated by the above, it will be of great significance for some high-dimensional dynamical systems with delays, such as secure communication \cite{DBLP:journals/mcs/BowongKK06} and image compression systems \cite{song2018multistability} to effectively synchronize the complex dynamic behaviors of these delayed systems in a fixed time. However, existing research results pay little attention to it. Therefore, this paper aims at is studied the FXTSNY of a class of UCQVMNNs with mixed delays by the non-decomposition method. The main contributions presented in this article can be described as follows.

\begin{itemize}
	\item{ 
		Most of the UCQVMNNs are decomposed into four RVMNNs by the decomposition method. The calculation and proof process has been carried out four times, which is quite complicated and increases the difficulty of theoretical derivation. Therefore, using the non-decomposition method of one-norm makes the proof result easier to realize and the calculation process more simple. In addition, the existence of memristor will cause the system to be discontinuous, few studies have used this approach for FXTSNY analysis of UCQVNNs with memristors.
	}
\end{itemize} 

\begin{itemize}
	\item{ 
		A novel FXTSNY method is applied to UCQVMNNs for the first time, combined with Lemma 2.2. Firstly, $V(t) \geq1$ or $0 < V(t) < 1$ is judged and the functioning part is intelligently chosen. Then, by using the improved one-norm, a suitable Lyapunov function and controller are constructed, and a more robust and accurate synchronization time estimation is obtained. 
	}
\end{itemize} 

\begin{itemize}
	\item{ 
		To verify the effectiveness of the non-decomposition method of the UCQVMNNs, several numerical simulations of the FXTSNY process are proposed. It not only verifies the correctness and validity of the two theorems, but also shows the superiority of Theorem 2. Moreover, we can achieve the best synchronization effect by adjusting the values of different parameters. Interestingly, UCQVMNNs have good applications in the rapid recovery of high-dimensional data, so it has practical research significance.
	}
\end{itemize}

The rest of this paper is organized as follows. Section 2 presents the description of QVMNNs with bilateral coefficients and unilateral coefficients, respectively. And several Definitions, Lemmas, and some new one-norm inequalities of the quaternion are deduced. In Section 3, the FXTSNY scheme is proposed while the effectiveness of the theoretical results is illustrated. The effectiveness of sufficient conditions is checked by simulation examples in Section 4. Some conclusions are drawn in Section 5.

\vspace{1ex}
\noindent $\bm{Notations}$: The sets of real and non-negative real numbers are represented by $\mathbb{R}$ and $\mathbb{R}^{+}$, respectively.
 $\mathbb{H}$ represents the set of all quaternions and denoted by bold letter, superscripts $T$ and $*$ indicate transposition and conjugate transpose, respectively. The set of all $n$-dimensional real numbers, complex numbers, and quaternions are represented by $\mathbb{R}^{n}$, $\mathbb{C}^{n}$ and $\mathbb{H}^{n}$. Using $\|\cdot\|_1$ and ${\rm sgn}(\cdot)$ to denote the one-norm and the sign function, respectively. $\mathbb{R}^{n\times m}$, $\mathbb{C}^{n\times m}$ and $\mathbb{H}^{n\times m}$ denote the $n\times m$-dimensional real, complex and quaternion matrix. Real and complex numbers are the special case of quaternions (i.e. $\mathbb{R}\in \mathbb{C} \in\mathbb{H}$).

 \section{Model Formulation and Preliminaries}
 
 \subsection{Quaternion Algebra Fundamentals}
 \label{}
A quaternion number $\bm x \in\mathbb{H}$ is combined by a real part and
three imagery parts, and can be written as following form:
\[\bm{x}=x_{(0)}+x_{(1)}i+x_{(2)}j+x_{(3)}k,\]
where $x_{(\zeta)}\in \mathbb{R}$, $\zeta =0,1,2,3$, and $i,j,k$ are  imaginary units, and the imaginary units are defined by
\begin{equation*}
	\begin{cases}
		i^2= j^2 =k^2= ijk=-1,\\
		ij=-ji = k,    jk=-kj = i,  ki=-ik = j.
	\end{cases}
\end{equation*}
The modulus of $\bm x$ are defined as $|\bm x|= \sqrt{\bm x\bar{\bm x}}
=\sqrt{(x_{(0)})^{2}+(x_{(1)})^{2}+(x_{(2)})^{2}+(x_{(3)})^{2}}$. The transpose of vector $\bm x$ is denoted by $x^{T}$, the conjugate and conjugate transpose of $\bm x$ are denoted by $\bar{\bm x}=x_{(0)}-x_{(1)}i-x_{(2)}j-x_{(3)}k$ and $\bm x^{\ast}=(x_{(0)}-x_{(1)}i-x_{(2)}j-x_{(3)}k)^{T}$, respectively. 
For any two quaternion $\bm x$ and $\bm y=y_{(0)}+iy_{(1)}+jy_{(2)}+ky_{(3)}$, the addition operation is defined as follows
\begin{equation}\nonumber 
	\bm x+\bm y=(x_{(0)}+y_{(0)})+(x_{(1)}+y_{(1)})i+(x_{(2)}+y_{(2)})j+(x_{(3)}+y_{(3)})k.
\end{equation}
The multiplication operation is defined by
\begin{equation}\nonumber
	\begin{aligned}
		\bm{xy}=& (x_{(0)}y_{(0)}-x_{(1)}y_{(1)}-x_{(2)}y_{(2)}-x_{(3)}y_{(3)})+(x_{(0)}y_{(1)}+x_{(1)}y_{(0)}+x_{(2)}y_{(3)}
		\\&-x_{(3)}y_{(2)})i
		+(x_{(0)}y_{(2)}-x_{(1)}y_{(3)}+x_{(2)}y_{(0)}+x_{(3)}y_{(1)})j
		+(x_{(0)}y_{(3)}
		\\&+x_{(1)}y_{(2)}-x_{(2)}y_{(1)}+x_{(3)}y_{(0)})k.
	\end{aligned}
\end{equation}
It is important to note that the multiplication in the quaternion domain is not commutative, i.e. $\bm{xy} \neq \bm{yx}.$

The one-norm of vector $v=(v_{1},v_{2},...,v_{n})\in \mathbb{R}^{n}$ and the quaternion $\bm x$ are written as $\|v\|_{1}=\sum_{p=1}^{n}|v_{p}|$ and $\|\bm x\|_{1}=\sum_{\zeta =0,1,2,3}\|x_{(\zeta)}\|_{1}$, respectively. For $\bm e(t)=(\bm e_{1}(t),\bm e_{2}(t),...,\bm e_{n}(t))^{T}\in \mathbb{H}^{n}$, $t\in \mathbb{R}$, the sign function and one-norm of vector $\bm e(t)$ are denoted by ${\rm sgn}(\bm e(t))=({\rm sgn}(\bm e_{1}(t)), {\rm sgn}(\bm e_{2}(t)),...,{\rm sgn}(\bm e_{n}(t)))^{T}$, $\|\bm e(t)\|_{1}=\sum_{p=1}^{n}\|\bm e_{p}(t)\|_{1}$ respectively, and $[\bm e(t)]^{r}=([\bm e_{1}(t)]^{r},[\bm e_{2}(t)]^{r},...,[\bm e_{n}(t)]^{r})^{T}$, moreover, 
$[\bm e_{p}(t)]^{r}={\rm sgn}(\bm e_{p}(t))\|\bm e_{p}(t)\|_{1}^{r}$, $p=1,2,...,n$ and $r>0$.

For any quaternion $\bm{x}=x_{(0)}+x_{(1)}i+x_{(2)}j+x_{(3)}k$ can be uniquely expressed as $\bm{x}=x_{1}+x_{2}j$, where $x_{1}=x_{(0)}+x_{(1)}i$, and $x_{2}=x_{(2)}+x_{(3)}i$. Furthermore, this expression can be used
by quaternion matrix.

\vspace{1ex}
\noindent \textbf{Definition 2.1} \cite{zhang1997quaternions}. Given a quaternion matrix $\bm A \in\mathbb{H}^{m\times n}$, its expression using the Cayley-Dickson notation is $\bm A= A_{p}+ A_{q}j$, where $ A_{p}$ and $ A_{q}\in\mathbb{C}^{m\times n}$. The quaternion matrix can be denoted as an isomorphic complex matrix
\begin{align*}
	&\bm A=\left(
	\begin{array}{cccccc}
		 A_{p}  &  A_{q}\\
		-\bar{ A_{q}} & \bar{ A_{p}}
	\end{array}
	\right),
\end{align*}
where $ A_{p}= A_{0}+ A_{1}i\in\mathbb{C}^{m\times n}$ and $ A_{q}= A_{2}+ A_{3}i\in\mathbb{C}^{m\times n}$.

\subsection{QVMNNs with Bilateral Coefficients (BCQVMNNs)}

   In this section, we introduce a new kind of bilateral coefficients QVMNNs (BCQVMNNs), which has discrete and distributed time delays and with a form	 
\begin{align}
	&\frac{d}{dt}\bm x_{p}(t)=
	- d_{p}\bm x_{p}(t)+\sum_{q=1}^{n} \grave{\bm a}_{pq}(\bm x_{p}(t))\bm f_{q}(\bm x_{q}(t))\acute{\bm a}_{pq}(\bm x_{p}(t)) \nonumber
	\\
	&\qquad\qquad \
	+\sum_{q=1}^{n} \grave{\bm b}_{pq}(\bm x_{p}(t))\bm g_{q}(\bm x_{q}(t-\tau(t)))\acute{\bm b}_{pq}(\bm x_{p}(t)) \nonumber
	\\
	&\qquad\qquad \
	+\sum_{q=1}^{n}\grave{\bm c}_{pq}(\bm x_{p}(t))(\int_{t-\pi}^{t}{\bm h_{q}(\bm x_{q}(s))}ds)\acute{\bm c}_{pq}(\bm x_{p}(t))+\bm I_{p},
	\label{qmc1}
\end{align}
for $p=1,2,...,n$, where $n$ corresponds to the number of neurons, $\bm x_{p}(t) \in \mathbb{H}$ stand the state variable of the $p$-th neuron, $d_{p}>0$ is the real-valued self-feedback coefficient, $\bm I_{p}\in \mathbb{H}$ denotes the external input or bias, $\grave{\bm a}_{pq}(\bm x_{p}(t)), \acute{\bm a}_{pq}(\bm x_{p}(t)), \grave{\bm b}_{pq}(\bm x_{p}(t))$, $\acute{\bm b}_{pq}(\bm x_{p}(t)), \grave{\bm c}_{pq}(\bm x_{p}(t)),$ and $\acute{\bm c}_{pq}(\bm x_{p}(t))\in \mathbb{H}$ stand for the memristive connection weights, $\bm f_{q}(\bm x_{q}(t)),$ $\bm g_{q}(\bm x_{q}(t-\tau(t)))$ and $\bm h_{q}(\bm x_{q}(s))\in \mathbb{H}$ are the activation functions. The discrete time delay and the distributed time delay are denoted by $\tau(t)$ and $\pi$, they satisfies $0 \leq\tau(t)\leq\tau, \tau={\rm max}\{\tau(t),\pi\}$. 

For simpler expression, we can represent \eqref{qmc1} as vector form
\begin{align}
	\frac{d}{dt}\bm{X}(t)=
	-D \bm{X}(t)+\grave{\bm A}\bm f(\bm{X}(t))\acute{\bm A} 
	+ \grave{\bm B}\bm g(\bm{X}(t-\tau(t)))\acute{\bm B} 
	+\grave{\bm C}(\int_{t-\pi}^{t}{\bm h(\bm{X}(s))}ds)\acute{\bm C}+\bm I,
	\label{qmc2}
\end{align}
where $\bm{X}(t)=(\bm x_{1}(t), \bm x_{2}(t),\cdots,\bm x_{n}(t))^{T}\in \mathbb{H}^{n}$, $D=diag(d_{1},d_{2},\cdots,d_{n})\in \mathbb{R}^{n\times n}$, $\bm f(\bm{X}(t))=(\bm f_{1}(\bm x_{1}(t)),\bm f_{2}(\bm x_{2}(t)),\cdots,\bm f_{n}(\bm x_{n}(t)))^{T}\in \mathbb{H}^{n}$, $\bm g(\bm{X}(t-\tau(t)))=(\bm g_{1}(\bm x_{1}(t-\tau(t))), \bm g_{2}(\bm x_{2}(t-\tau(t))), \cdots, \bm g_{n}(\bm x_{n}(t-\tau(t))))^{T}\in \mathbb{H}^{n}$, 
$\bm h(\bm{X}(s))=(\bm h_{1}(\bm x_{1}(s)), \bm h_{2}(\bm x_{2}(s)),$ $\cdots, \bm h_{n}(\bm x_{n}(s)))^{T}\in \mathbb{H}^{n}$, $\bm I=(\bm I_{1}, \bm I_{2}, \cdots, \bm I_{n})^{T}\in \mathbb{H}^{n}$ and 
\begin{equation}   \nonumber
\grave{\bm A}=\left(                
	\begin{array}{cccc}
		\grave{\bm a}_{11}(\bm x_{1}(t)) & \grave{\bm a}_{12}(\bm x_{1}(t)) &\cdots& \grave{\bm a}_{1n}(\bm x_{1}(t))\\  		
		\grave{\bm a}_{21}(\bm x_{2}(t)) & \grave{\bm a}_{22}(\bm x_{2}(t)) & \cdots & \grave{\bm a}_{2n}(\bm x_{2}(t))\\  
		\vdots & \vdots & \ddots &\vdots\\ 	
		\grave{\bm a}_{n1}(\bm x_{n}(t)) & \grave{\bm a}_{n2}(\bm x_{n}(t)) & \cdots & \grave{\bm a}_{nn}(\bm x_{n}(t))\\ 	
	\end{array}	
	\right) \in \mathbb{H}^{n\times n},              
\end{equation}
\begin{equation}  \nonumber     	
	\acute{\bm A}=\left(                 	
	\begin{array}{cccc}
		\acute{\bm a}_{11}(\bm x_{1}(t)) & \acute{\bm a}_{12}(\bm x_{1}(t)) &\cdots& \acute{\bm a}_{1n}(\bm x_{1}(t))\\  		
		\acute{\bm a}_{21}(\bm x_{2}(t)) & \acute{\bm a}_{22}(\bm x_{2}(t)) & \cdots & \acute{\bm a}_{2n}(\bm x_{2}(t))\\  
		\vdots & \vdots & \ddots &\vdots\\ 	
		\acute{\bm a}_{n1}(\bm x_{n}(t)) & \acute{\bm a}_{n2}(\bm x_{n}(t)) & \cdots & \acute{\bm a}_{nn}(\bm x_{n}(t))\\ 	
	\end{array}	
	\right) \in \mathbb{H}^{n\times n},              	
\end{equation}
\begin{equation}   \nonumber
	\grave{\bm B}=\left(                
	\begin{array}{cccc}
		\grave{\bm b}_{11}(\bm x_{1}(t)) & \grave{\bm b}_{12}(\bm x_{1}(t)) &\cdots& \grave{\bm b}_{1n}(\bm x_{1}(t))\\  		
		\grave{\bm b}_{21}(\bm x_{2}(t)) & \grave{\bm b}_{22}(\bm x_{2}(t)) & \cdots & \grave{\bm b}_{2n}(\bm x_{2}(t))\\  
		\vdots & \vdots & \ddots &\vdots\\ 	
		\grave{\bm b}_{n1}(\bm x_{n}(t)) & \grave{\bm b}_{n2}(\bm x_{n}(t)) & \cdots & \grave{\bm b}_{nn}(\bm x_{n}(t))\\ 	
	\end{array}	
	\right) \in \mathbb{H}^{n\times n},              
\end{equation}
\begin{equation}  \nonumber     
	\acute{\bm B}=\left(              
	\begin{array}{cccc}
		\acute{\bm b}_{11}(\bm x_{1}(t)) & \acute{\bm b}_{12}(\bm x_{1}(t)) &\cdots& \acute{\bm b}_{1n}(\bm x_{1}(t))\\  		
		\acute{\bm b}_{21}(\bm x_{2}(t)) & \acute{\bm b}_{22}(\bm x_{2}(t)) & \cdots & \acute{\bm b}_{2n}(\bm x_{2}(t))\\  
		\vdots & \vdots & \ddots &\vdots\\ 	
		\acute{\bm b}_{n1}(\bm x_{n}(t)) & \acute{\bm b}_{n2}(\bm x_{n}(t)) & \cdots & \acute{\bm b}_{nn}(\bm x_{n}(t))\\ 	
	\end{array}	
	\right)\in \mathbb{H}^{n\times n},                 
\end{equation}
\begin{equation}   \nonumber
	\grave{\bm C}=\left(                
	\begin{array}{cccc}
		\grave{\bm c}_{11}(\bm x_{1}(t)) & \grave{\bm c}_{12}(\bm x_{1}(t)) &\cdots& \grave{\bm c}_{1n}(\bm x_{1}(t))\\  		
		\grave{\bm c}_{21}(\bm x_{2}(t)) & \grave{\bm c}_{22}(\bm x_{2}(t)) & \cdots & \grave{\bm c}_{2n}(\bm x_{2}(t))\\  
		\vdots & \vdots & \ddots &\vdots\\ 	
		\grave{\bm c}_{n1}(\bm x_{n}(t)) & \grave{\bm c}_{n2}(\bm x_{n}(t)) & \cdots & \grave{\bm c}_{nn}(\bm x_{n}(t))\\ 	
	\end{array}	
	\right) \in \mathbb{H}^{n\times n},              
\end{equation}
\begin{equation}  \nonumber     
	\acute{\bm C}=\left(             
	\begin{array}{cccc}
		\acute{\bm c}_{11}(\bm x_{1}(t)) & \acute{\bm c}_{12}(\bm x_{1}(t)) &\cdots& \acute{\bm c}_{1n}(\bm x_{1}(t))\\  		
		\acute{\bm c}_{21}(\bm x_{2}(t)) & \acute{\bm c}_{22}(\bm x_{2}(t)) & \cdots & \acute{\bm c}_{2n}(\bm x_{2}(t))\\  
		\vdots & \vdots & \ddots &\vdots\\ 	
		\acute{\bm c}_{n1}(\bm x_{n}(t)) & \acute{\bm c}_{n2}(\bm x_{n}(t)) & \cdots & \acute{\bm c}_{nn}(\bm x_{n}(t))\\ 	
	\end{array}	
	\right) \in \mathbb{H}^{n\times n}.               
\end{equation}
According to Definition 2.1, let $\bm{X}(t)=X_{1}(t)+X_{2}(t)j$, $\grave{\bm A}=\grave{A}_{1}+\grave{A}_{2}j$, $\acute{\bm A}=\acute{A}_{1}+\acute{A}_{2}j$, $\grave{\bm B}=\grave{B}_{1}+\grave{B}_{2}j$, $\acute{\bm B}=\acute{B}_{1}+\acute{B}_{2}j$, $\grave{\bm C}=\grave{C}_{1}+\grave{C}_{2}j$, $\acute{\bm C}=\acute{C}_{1}+\acute{C}_{2}j$, and $\bm f(\bm{X}(t))=f_{1}(\bm{X}(t))+f_{2}(\bm{X}(t))j$, 
$\bm g(\bm{X}(t-\tau(t)))=g_{1}(\bm{X}(t-\tau(t)))+g_{2}(\bm{X}(t-\tau(t)))j$, $\bm h(\bm{X}(s))=h_{1}(\bm{X}(s))+h_{2}(\bm{X}(s))j$, $\bm I=L_{1}+L_{2}j$, where $X_{1}(t)$, $X_{2}(t),$ $L_{1}, L_{2}$, $f_{1}(\bm{X}(t))$, $f_{2}(\bm{X}(t))$, $g_{1}(\bm{X}(t-\tau(t))), g_{2}(\bm{X}(t-\tau(t)))$, and $h_{1}(\bm{X}(s)), h_{2}(\bm{X}(s)) \in\mathbb{C}^{n}$, $\grave{A}_{1}, \grave{A}_{2}$, $\acute{A}_{1}, \acute{A}_{2}$, $\grave{B}_{1}, \grave{B}_{2}$, $\acute{B}_{1}, \acute{B}_{2}$, $\grave{C}_{1}, \grave{C}_{2}$, and $\acute{C}_{1}, \acute{C}_{2} \in \mathbb{C}^{n\times n} $.

Then system \eqref{qmc2} can be converted to the following form
\begin{align*}
&	\frac{d}{dt}(X_{1}(t)+X_{2}(t)j)=
	-D (X_{1}(t)+X_{2}(t)j)+(\grave{A}_{1}+\grave{A}_{2}j) (f_{1}(\bm{X}(t))+f_{2}(\bm{X}(t))j)(\acute{A}_{1}+\acute{A}_{2}j ) 
	\\
	& \qquad \qquad \qquad \qquad \quad \
	+(\grave{B}_{1}+\grave{B}_{2}j) \bm (g_{1}(\bm{X}(t-\tau(t)))+g_{2}(\bm{X}(t-\tau(t)))j)(\acute{B}_{1}+\acute{B}_{2}j)
	\\
	& \qquad \qquad \qquad \qquad \quad \
	+(\grave{C}_{1}+\grave{C}_{2}j)(\int_{t-\pi}^{t}{(h_{1}(\bm{X}(s))+h_{2}(\bm{X}(s))j)}ds)(\acute{C}_{1}+\acute{C}_{2}j)
	\\
	& \qquad \qquad \qquad \qquad \quad \
	+L_{1}+L_{2}j.
\end{align*}
Thus system \eqref{qmc2} is equivalent to
\begin{align}
	&\frac{d}{dt}X_{1}(t)=
	-D X_{1}(t)
	+\grave{A}_{1}\acute{A}_{1}f_{1}(\bm{X}(t))-\grave{A}_{1}\bar{\acute{A}}_{2} f_{2}(\bm{X}(t))-\grave{A}_{2}\bar{\acute{A}}_{2}\bar{f_{1}}(\bm{X}(t))-\grave{A}_{2}\acute{A}_{1} \bar{f_{2}}(\bm{X}(t)) \nonumber
	\\
	& \qquad \qquad \ \
	+\grave{B}_{1}\acute{B}_{1}g_{1}(\bm{X}(t-\tau(t)))-\grave{B}_{1}\bar{\acute{B}}_{2} g_{2}(\bm{X}(t-\tau(t)))-\grave{B}_{2}\bar{\acute{B}}_{2}\bar{g_{1}}(\bm{X}(t-\tau(t))) \nonumber
	\\
	& \qquad \qquad \ \
	-\grave{B}_{2}\acute{B}_{1} \bar{g}_{2}(\bm{X}(t-\tau(t)))
	+\grave{C}_{1}\acute{C}_{1}\int_{t-\pi}^{t}{(h_{1}(\bm{X}(s))}ds-\grave{C}_{1}\bar{\acute{C}}_{2}\int_{t-\pi}^{t}{h_{2}(\bm{X}(s))}ds \nonumber
	\\
	& \qquad \qquad \ \
	-\grave{C}_{2}\bar{\acute{C}}_{2}\int_{t-\pi}^{t}{(\bar{h}_{1}(\bm{X}(s))}ds-\grave{C}_{2}\acute{C}_{1}\int_{t-\pi}^{t}{\bar{h}_{2}(\bm{X}(s))}
	+L_{1},
	\label{qmc3}
\end{align}
and
\begin{align}
	&\frac{d}{dt}X_{2}(t)=
	-DX_{2}(t)
	+\grave{A}_{1}\acute{A}_{2}f_{1}(\bm{X}(t))+\grave{A}_{1}\bar{\acute{A}}_{1} f_{2}(\bm{X}(t))+\grave{A}_{2}\bar{\acute{A}}_{1}\bar{f}_{1}(\bm{X}(t))-\grave{A}_{2}\acute{A}_{2} \bar{f_{2}}(\bm{X}(t)) \nonumber
	\\
	& \qquad \qquad \
	+\grave{B}_{1}\acute{B}_{2}g_{1}(\bm{X}(t-\tau(t)))+\grave{B}_{1}\bar{\acute{B}}_{1} g_{2}(\bm{X}(t-\tau(t)))+\grave{B}_{2}\bar{\acute{B}}_{1}\bar{g}_{1}(\bm{X}(t-\tau(t))) \nonumber
	\\
	& \qquad \qquad \
	-\grave{B}_{2}\acute{B}_{2} \bar{g}_{2}(\bm{X}(t-\tau(t)))
	+\grave{C}_{1}\acute{C}_{2}\int_{t-\pi}^{t}{(h_{1}(\bm{X}(s))}ds
	+\grave{C}_{1}\bar{\acute{C}}_{1}\int_{t-\pi}^{t}{h_{2}(\bm{X}(s))}ds \nonumber
	\\
	& \qquad \qquad \
	+\grave{C}_{2}\bar{\acute{C}}_{1}\int_{t-\pi}^{t}{(\bar{h}_{1}(\bm{X}(s))}ds
	-\grave{C}_{2}\acute{C}_{2}\int_{t-\pi}^{t}{\bar{h}_{2}(\bm{X}(s))}
	+L_{2}.
	\label{qmc4}
\end{align}

By the Cayley-Dickson transformation, BCQVMNNs \eqref{qmc2} is transformed into two CVMNNs \eqref{qmc3} and \eqref{qmc4}. Due to the mixed time delay and integral term in BCQVMNNs, it is difficult to directly study the FXTSYN of BCQVMNNs by non-decomposition method. As a special case of BCQVMNNs, the following part of this paper mainly illustrates the related properties of the bilateral systems by studying the sufficient conditions for FXTSYN of UCQVMNNs.

\subsection{QVMNNs with Unilateral Coefficients (UCQVMNNs)}
 
The following UCQVMNNs with discrete and distributed time delays is considered 
\begin{align}
	&\frac{d}{dt}\bm x_{p}(t)=
	-d_{p}\bm x_{p}(t)+\sum_{q=1}^{n} \bm a_{pq}(\bm x_{p}(t))\bm f_{q}(\bm x_{q}(t))
	+\sum_{q=1}^{n} \bm b_{pq}(\bm x_{p}(t))\bm g_{q}(\bm x_{q}(t-\tau(t))) \nonumber
	\\
	&\qquad\quad \
	+\sum_{q=1}^{n}\bm c_{pq}(\bm x_{p}(t))\int_{t-\pi}^{t}{\bm h_{q}(\bm x_{q}(s))}ds+\bm I_{p},
	\label{2.1}
\end{align}
for $p=1,2,...,n$, $\bm x_{p}(t) \in \mathbb{H}$ is the state variable, 
$\bm a_{pq}(\bm x_{p}(t))$, $\bm b_{pq}(\bm x_{p}(t))$, $\bm c_{pq}(\bm x_{p}(t))\in \mathbb{H}$ stand for the memristive connection weights.

Define system \eqref{2.1} as drive system, the response system is described as following
\begin{align}
	&\frac{d}{dt}\bm y_{p}(t)=
	-d_{p}\bm y_{p}(t)+\sum_{q=1}^{n} \bm a_{pq}(\bm y_{p}(t))\bm f_{q}(\bm y_{q}(t))
	+\sum_{q=1}^{n}\bm b_{pq}(\bm y_{p}(t))\bm g_{q}(\bm y_{q}(t-\tau(t)))\nonumber
	\\
	&\qquad\quad \
	+\sum_{q=1}^{n}\bm c_{pq}(\bm y_{p}(t))\int_{t-\pi}^{t}{\bm h_{q}(\bm y_{q}(s))}ds+\bm I_{p}+\bm u_{p}(t),
	\label{2.2}
\end{align}
where $\bm u_{p}(t)\in \mathbb{H}$ is the designed controller. The initial conditions of systems \eqref{2.1} and \eqref{2.2} are
\[
\bm x_{p}(s)=\bm \phi_{p}(s), \quad \bm y_{p}(s)=\bm \psi_{p}(s), s\in [-\tau,0]. 
\]

Based on the characteristics of the memristor and current–voltage, the memristive connection weights in \eqref{2.1} and \eqref{2.2} are satisfy
\begin{align*}
	&\bm a_{pq}(\cdot)=
	\begin{cases}
		\hat{\bm a}_{pq},\quad|\cdot|\leq r_{p},
		\\[1mm]
		\check{\bm a}_{pq},\quad|\cdot|>r_{p},
	\end{cases}
 \hspace{-0.05in}
	\bm b_{pq}(\cdot)=
	\begin{cases}
		\hat{\bm b}_{pq},\quad|\cdot|\leq r_{p},
		\\[1mm]
		\check{\bm b}_{pq},\quad|\cdot|>r_{p},
	\end{cases}
\hspace{-0.05in}
	\bm c_{pq}(\cdot)=
	\begin{cases}
		\hat{\bm c}_{pq},\quad|\cdot|\leq r_{p},
		\\[1mm]
		\check{\bm c}_{pq},\quad|\cdot|>r_{p},
	\end{cases}
\end{align*}

\noindent for $p,q=1,2,...,n$, where $\hat{\bm a}_{pq}$, $\check{\bm a}_{pq}$, $\hat{\bm b}_{pq}$,
$\check{\bm b}_{pq}$, $\hat{\bm c}_{pq}$,
$\check{\bm c}_{pq}$ are known constants with respect to the memristor, and the switching jumps $r_{p}>0$ is the threshold level. Define $a^{+}_{pq}={\rm max}\{|\hat{\bm a}_{pq}|,|\check{\bm a}_{pq}|\}$, $b^{+}_{pq}={\rm max}\{|\hat{\bm b}_{pq}|,|\check{\bm b}_{pq}|\}$ and $c^{+}_{pq}={\rm max}\{|\hat{\bm c}_{pq}|,|\check{\bm c}_{pq}|\}$. 
$\bm a^{\rm max}_{pq}={\rm max}\{\hat{\bm a}_{pq},\check{\bm a}_{pq}\}$, 
$\bm a^{\rm min}_{pq}={\rm min}\{\hat{\bm a}_{pq},\check{\bm a}_{pq}\}$, 
$\bm b^{\rm max}_{pq}={\rm max}\{\hat{\bm b}_{pq},\check{\bm b}_{pq}\}$, 
$\bm b^{\rm min}_{pq}={\rm min}\{\hat{\bm b}_{pq},\check{\bm b}_{pq}\}$, 
$\bm c^{\rm max}_{pq}={\rm max}\{\hat{\bm c}_{pq},\check{\bm c}_{pq}\}$, and 
$\bm c^{\rm min}_{pq}={\rm min}\{\hat{\bm c}_{pq},\check{\bm c}_{pq}\}$.

As shown above, due to the discontinuity of the memristive connection weights, the system \eqref{2.1} and the system \eqref{2.2} are regarded as the discontinuous differential equation of the right-hand side, which the existence and uniqueness of solutions are not guaranteed. So the Filippov solutions are utilized to deal with the special case.

By the differential inclusions and the set-valued maps \cite{filippov2013differential}, the differential inclusion of systems \eqref{2.1} and \eqref{2.2} are as follows
\begin{align}
	&\frac{d}{dt}\bm x_{p}(t)\in
	-d_{p}\bm x_{p}(t)+\sum_{q=1}^{n}\bm K(\bm a_{pq}(\bm x_{p}(t)))\bm f_{q}(\bm x_{q}(t))
	+\sum_{q=1}^{n}\bm K(\bm b_{pq}(\bm x_{p}(t))) \nonumber
	\\
	&\qquad \qquad \
	\times\bm g_{q}(\bm x_{q}(t-\tau(t)))+\sum_{q=1}^{n}\bm K(\bm c_{pq}(\bm x_{p}(t)))\int_{t-\pi}^{t}{\bm h_{q}(\bm x_{q}(s))}ds+\bm I_{p},
	\label{2.3}
\end{align}
and
\begin{align}
	&\frac{d}{dt}\bm y_{p}(t)\in
	-d_{p}\bm y_{p}(t)+\sum_{q=1}^{n}\bm K(\bm a_{pq}(\bm y_{p}(t)))\bm f_{q}(\bm y_{q}(t))
	+\sum_{q=1}^{n}\bm K(\bm b_{pq}(\bm y_{p}(t)))\nonumber
	\\
	&\qquad \qquad \
	\times\bm g_{q}(\bm y_{q}(t-\tau(t)))+\sum_{q=1}^{n}\bm K(\bm c_{pq}(\bm y_{p}(t)))\int_{t-\pi}^{t}{\bm h_{q}(\bm y_{q}(s))}ds+\bm I_{p}+\bm u_{p}(t),
	\label{2.4}
\end{align}
where
\begin{align*}
	&\bm K(\bm a_{pq}(\cdot))=\left\{
	\begin{aligned}
		\hat{\bm a}_{pq},\quad\ |\cdot|< r_{p},
		\\
		{\rm co}\{\hat{\bm a}_{pq},\check{\bm a}_{pq}\},\
		|\cdot|= r_{p},
		\\
		\check{\bm a}_{pq},\quad|\cdot|>r_{p},
	\end{aligned}
	\right.
	\qquad
	\bm K(\bm b_{pq}(\cdot))=\left\{
	\begin{aligned}
		\hat{\bm b}_{pq},\quad|\cdot|< r_{p},
		\\
		{\rm co}\{\hat{\bm b}_{pq},\check{\bm b}_{pq}\},\
		|\cdot|= r_{p},
		\\
		\check{\bm b}_{pq},\quad|\cdot|>r_{p},
	\end{aligned}
	\right.
\end{align*}
\begin{align*}
	&\bm K(\bm c_{pq}(\cdot))=\left\{
	\begin{aligned}
		\hat{\bm c}_{pq},\quad\ |\cdot|< r_{p},
		\\
		{\rm co}\{\hat{\bm c}_{pq},\check{\bm c}_{pq}\},\
		|\cdot|= r_{p},
		\\
		\check{\bm c}_{pq},\quad|\cdot|>r_{p},
	\end{aligned}
	\right.
\end{align*} 
in which ${\rm co}\{a,b\}$ is the closure of the convex hull, and ${\rm co}\{\hat{\bm a}_{pq},\check{\bm a}_{pq}\}=[\bm a^{\rm min}_{pq}, \bm a^{\rm max}_{pq}]$, ${\rm co}\{\hat{\bm b}_{pq},\check{\bm b}_{pq}\}=[\bm b^{\rm min}_{pq}, \bm b^{\rm max}_{pq}]$,  ${\rm co}\{\hat{\bm c}_{pq},\check{\bm c}_{pq}\}=[\bm c^{\rm min}_{pq}, \bm c^{\rm max}_{pq}]$.

By the measurable selection theorem \cite{clarke1990optimization}, if $(\bm x_{p}(t),\bm y_{p}(t))$ is the solution of systems \eqref{2.3}-\eqref{2.4}, there exist bounded measurable functions $\tilde{\bm a}_{pq}(t)\in \bm K(\bm a_{pq}(\cdot))$, $\tilde{\bm b}_{pq}(t)\in \bm K(\bm b_{pq}(\cdot))$ and $\tilde{\bm c}_{pq}(t)\in \bm K(\bm c_{pq}(\cdot))$, which rely on $\bm x_{p}(t)$ and $\bm y_{p}(t)$ respectively, such that
\begin{align}
	&\frac{d}{dt}\bm x_{p}(t)=
	-d_{p}\bm x_{p}(t)+\sum_{q=1}^{n}\tilde{\bm a}_{pq}(t)\bm f_{q}(\bm x_{q}(t))
	+\sum_{q=1}^{n}\tilde{\bm b}_{pq}(t) \nonumber
	\\
	&\qquad\quad \
	\times\bm g_{q}(\bm x_{q}(t-\tau(t)))+\sum_{q=1}^{n}\tilde{\bm c}_{pq}(t)\int_{t-\pi}^{t}{\bm h_{q}(\bm x_{q}(s))}ds+\bm I_{p},
	\label{2.5}
\end{align}
and
\begin{align}
	&\frac{d}{dt}\bm y_{p}(t)=
	-d_{p}\bm y_{p}(t)+\sum_{q=1}^{n}\tilde{\bm a}_{pq}(t)\bm f_{q}(\bm y_{q}(t))
	+\sum_{q=1}^{n}\tilde{\bm b}_{pq}(t)\bm g_{q}(\bm y_{q}(t-\tau(t)))\nonumber
	\\
	&\qquad\quad \
	+\sum_{q=1}^{n}\tilde{\bm c}_{pq}(t)\int_{t-\pi}^{t}{\bm h_{q}(\bm y_{q}(s))}ds+\bm I_{p}+\bm u_{p}(t).
	\label{2.6}
\end{align}
Based on the UCQVMNNs \eqref{2.5} and \eqref{2.6}, the synchronization error can be defined as $\bm e_{p}(t)=\bm y_{p}(t)-\bm x_{p}(t)$, so we can get
\begin{align}
	&\frac{d}{dt}\bm e_{p}(t)=
	-d_{p}\bm e_{p}(t)+\sum_{q=1}^{n}\tilde{\bm a}_{pq}(t)(\bm f_{q}(\bm y_{q}(t)-\bm f_{q}(\bm x_{q}(t)))\nonumber
	\\
	&\qquad\quad \
	+\sum_{q=1}^{n}\tilde{\bm b}_{pq}(t)(\bm g_{q}(\bm y_{q}(t-\tau(t)))-\bm g_{q}(\bm x_{q}(t-\tau(t))))\nonumber
	\\
	&\qquad\quad \
	+\sum_{q=1}^{n}\tilde{\bm c}_{pq}(t)\int_{t-\pi}^{t}{(\bm h_{q}(\bm y_{q}(s))-\bm h_{q}(\bm x_{q}(s)))}ds+\bm u_{p}(t),
	\label{2.7}
\end{align}
with the initial condition $\bm \varphi_{p}(s)=\bm \psi_{p}(s)-\bm \phi_{p}(s)$ for $s\in[-\tau,0]$.

To obtain the main results, the following Assumptions, Definitions and Lemmas are necessary.

\vspace{2ex}
\noindent{\bf Assumption (A1)}. 
For any $\bm x_{q}(t), \bm y_{q}(t)\in \mathbb{H}$, suppose the expression of activation functions 
\[ \bm f_{q}(\cdot)=f_{q(0)}(\cdot)+i f_{q(1)}(\cdot)
+jf_{q(2)}(\cdot)+kf_{q(3)}(\cdot),
\] 
\[\bm g_{q}(\cdot)=g_{q(0)}(\cdot)+i g_{q(1)}(\cdot)+jg_{q(2)}(\cdot)+k g_{q(3)}(\cdot),\] 
\[\bm h_{q}(\cdot)=h_{q(0)}(\cdot)+i h_{q(1)}(\cdot)+jh_{q(2)}(\cdot)+k f_{q(3)}(\cdot).\] 
where $\bm f_{q}(0)=0, \bm g_{q}(0)=0, \bm h_{q}(0)=0$, $f_{q(\zeta)}(\cdot), g_{q(\zeta)}(\cdot), h_{q(\zeta)}(\cdot)\in \mathbb{R}$, 
$\zeta=0,1,2,3$, 
then there exist positive constant $\upsilon_{q}, \varrho_{q}$ and $\iota_{q}, q=1,2,...,n$ satisfying
\begin{align*}
	&\|\bm f_{q}(\bm y_{q}(t)-\bm f_{q}(\bm x_{q}(t))\|_{1}\leq \upsilon_{q}\|\bm y_{q}(t)-\bm x_{q}(t)\|_{1}=\upsilon_{q}\|\bm e_{q}(t)\|_{1},
	\\
	&
	\|\bm g_{q}(\bm y_{q}(t)-\bm g_{q}(\bm x_{q}(t))\|_{1}\leq \varrho_{q}\|\bm y_{q}(t)-\bm x_{q}(t)\|_{1}=\varrho_{q}\|\bm e_{q}(t)\|_{1},
	\\  
	&
	\|\bm h_{q}(\bm y_{q}(t)-\bm h_{q}(\bm x_{q}(t))\|_{1}\leq \iota_{q}\|\bm y_{q}(t)-\bm x_{q}(t)\|_{1}=\iota_{q}\|\bm e_{q}(t)\|_{1}.
\end{align*}

\vspace{1ex}

\noindent \textbf{Definition 2.2} \cite{chen2017fixed}. The drive system \eqref{2.1} is said to be synchronized with the response system \eqref{2.2} in finite time, if there exists a constant $T(e_{0})>0$ and $T(e_{0})$ depends on the initial error $e_{0}$ such that
\[
\lim_{t\rightarrow T (e_{0})}\|e(t)\|=0, \qquad
\|e(t)\|=0 \quad {\rm for}\ \ \forall t\geq T(e_{0}),
\]
where $e(t)=(e_{1}(t),e_{1}(t),\cdots e_{n}(t))^{T}$, and $T(e_{0})$ is called the settling time.
\vspace{2ex}

\noindent \textbf{Definition 2.3} \cite{polyakov2011nonlinear}.
The drive system \eqref{2.1} and the response system \eqref{2.2} are said to reach the fixed-time synchronization, if the following two conditions hold

(1)\ The system \eqref{2.1} is said to be synchronized with the system \eqref{2.2} in finite time,

(2)\ There exists a fixed constant $T_{max}$, such that for any initial condition $e_{0}$, the corresponding settling time satisfies $T(e_{0})\leq T_{max}$.
\vspace{2ex}

\noindent \textbf{Lemma 2.1} \cite{polyakov2011nonlinear}. Assume that there exists a continuous radically unbounded function $V:\mathbb{R}^{n}\rightarrow \mathbb{R}^{+}=[0,+\infty)$ satisfying

(1)\ $V(e(t))=0 $ if and only if $e(t)=0$,

(2)\ $\frac{d}{dt}V(e(t))
\leq {-aV^{\alpha}(e(t))}-{bV^{\beta}(e(t))}$,

\noindent where $a,b>0$, $0<\alpha<1$, and $\beta>1$.

Then the origin of the error system \eqref{2.7} is fixed-time stable, and $V(e(t))=0$ for $t\geq T(e_{0})$ where the settling time $T(e_{0})$ is bounded by
\[
T(e_{0})\leq T_{max}=
\frac{1}{a}\left(\frac{a}{b}\right)^{\frac{1-\alpha}{\beta-\alpha}}\left(\frac{1}{\beta-1}+\frac{1}{1-\alpha}\right)\qquad {\rm for}\ \ \forall\ e_{0}\in \mathbb{R}^{n}.
\]

\begin{remark} {\rm\cite{polyakov2011nonlinear}}.
	Under the same conditions in Lemma 2.1, the synchronization time $T(e_{0})$ can also be estimated by the following formula 
	\[
	T(e_{0})\leq T_{max}=
	\frac{1}{a(1-\alpha)}+\frac{1}{b(\beta-1)} \qquad {\rm for}\ \  \forall e_{0}\in \mathbb{R}^{n}.
	\]
\end{remark}

\noindent \textbf{Lemma 2.2} \cite{li2020fixed}.
If there exists a continuous, positive definite, and radically unbounded function $V(e(t)):\mathbb{R}^{n}\rightarrow \mathbb{R}^{+}$ such that any solution $x(t)$ of system \eqref{2.7} satisfies the inequality
\begin{align}
	\frac{d}{dt}V(e(t))\leq \left\{
	\begin{aligned}
		\quad	{aV(e(t))}-{b_{1}(V(e(t)))^{\gamma+{\rm sgn}(V(e(t))-1)}},\qquad \quad  V\geqslant 1,
		\\
		{aV(e(t))}-{b_{2}(V(e(t)))^{\gamma+{\rm sgn}(V(e(t))-1)}},\quad 0\leq V<1,
	\end{aligned}
	\right.
\end{align} 
in which $a<\min\{b_{1},b_{2}\}$, $b_{1},b_{2}>0$, $1\leq \gamma<2$, then the origin of the error system \eqref{2.7} is globally fixed-time stable, In addition, for any initial state $e_{0}$ of system \eqref{2.7}, the settling time is described as
\begin{align}
	&T=\left\{
	\begin{aligned}
		\frac{1}{a(2-\gamma)}\ln\frac{b_{2}}{b_{2}-a}+\frac{1}{\gamma(b_{1}-a)},\quad\ a>0,
		\\
		\frac{1}{b_{2}(2-\gamma)}+\frac{1}{b_{1}\gamma},\ \qquad \qquad \qquad \quad a=0,
		\\
		\quad	\frac{1}{a(2-\gamma)}\ln\frac{b_{2}}{b_{2}-a}+\frac{1}{a\gamma}\ln\frac{b_{1}}{b_{1}-a},\quad a<0,
	\end{aligned}
	\right.
	\label{2.9}
\end{align} 
\vspace{1ex}

\noindent \textbf{Lemma 2.3} \cite{hardy1952inequalities}. If $x_{1},x_{2},\cdots,x_{n}\geq 0$, $0<q_1\leq 1$ and $p_1>1$, then
\[
\sum_{p=1}^{n} x_{p}^{q_1}\geq\left (\sum_{p=1}^{n} x_{p}\right)^{q_1},\qquad
\sum_{p=1}^{n} x_{p}^{p_1}\geq n^{1-p_1}\left(\sum_{p=1}^{n} x_{p}\right)^{p_1}.
\]

\noindent \textbf{Definition 2.4} \cite{deng2019fixed}.
We define sign function of quaternion $\bm x=x_{(0)}+x_{(1)}i+x_{(2)}j+x_{(3)}k$ as follows
\[{\rm sgn}(\bm x)={\rm sgn}(x_{(0)})+{\rm sgn}(x_{(1)})i+{\rm sgn}(x_{(2)})j+{\rm sgn}(x_{(3)})k,\]
and the conjugate transpose of ${\rm sgn}(x)$ are denoted by
\[{\rm sgn}(\bm x)^{*}=\big({\rm sgn}(x_{(0)})-{\rm sgn}(x_{(1)})i-{\rm sgn}(x_{(2)})j-{\rm sgn}(x_{(3)})k\big)^{T}.\]

\noindent \textbf{Definition 2.5.}
According to the characteristics of the sign function, it can be known that $|x|={\rm sgn}(x)x,  x\in \mathbb{R}$, and the one-norm of quaternoin vector $\bm u=u_{(0)}+u_{(1)}i+u_{(2)}j+u_{(3)}k \in \mathbb{H}^{n}$ can be expressed as
\begin{equation}\nonumber
	\begin{aligned}
		\|\bm u\|_{1}&= \|u_{(0)}\|_{1}+\|u_{(1)}\|_{1}+\|u_{(2)}\|_{1}+\|u_{(3)}\|_{1}
		\\&=({\rm sgn}(u_{(0)}))^{T}u_{(0)}+({\rm sgn}(u_{(1)}))^{T}u_{(1)}+({\rm sgn}(u_{(2)}))^{T}u_{(2)}+({\rm sgn}(u_{(3)}))^{T}u_{(3)}
		\\&=\frac{1}{2}(({\rm sgn}(\bm u))^{*}\bm u+{\rm sgn}(\bm u)\bm u^{*}).
	\end{aligned}
\end{equation}
where $u_{(\zeta)}\in \mathbb{R}^{n},\zeta=0,1,2,3$ represent the real and imaginary parts of the quaternoin vector, respectively.

There are some Lemmas associated with this Definition is given below, so as to facilitate the proof and calculation of the theorem later.
\vspace{1ex}

\noindent \textbf{Lemma 2.4} \cite{peng2022finite}.
Suppose that vector $\bm e(t)=(\bm e_{1}(t),\bm e_{2}(t),...,\bm e_{n}(t))^{T}$, 
$\bm l(t)=(\bm l_{1}(t),\bm l_{2}(t) \\
,...,\bm l_{n}(t))^{T}\in \mathbb{H}^{n}$, and $\bm e(t)=e_{(0)}(t)+e_{(1)}(t)i+e_{(2)}(t)j+e_{(3)}(t)k$,
$\bm l(t)=l_{(0)}(t)+l_{(1)}(t)i+l_{(2)}(t)j+l_{(3)}(t)k$, where $\bm e_{p}(t), \bm l_{p}(t)\in \mathbb{H},$ $ p=1,2,...,n$, and $e_{(\zeta)}(t), l_{(\zeta)}(t)\in \mathbb{R}^{n}, \zeta =0,1,2,3$. For any $\bm e(t)$, $\bm l(t)\in \mathbb{H}^{n}$, positive constant $c>0$, if $\bm f(\bm e(t))$ is an integrable function is defined on $[t-\tau,t]$, the following formulas hold
\begin{flalign*}
	&(1)\ {\rm sgn}(\bm e(t))^{*}\bm e(t)+(\bm e(t))^{*}{\rm sgn}(\bm e(t))=2\|\bm e(t)\|_{1},
	\\
	&(2)\ {\rm sgn}(\bm e(t))^{*}\bm l(t)+(\bm l(t))^{*}{\rm sgn}(\bm e(t))\leq 2\|\bm l(t)\|_{1},
	\\
	&(3)\ {\rm sgn}(\bm e(t))^{*}{\rm sgn}(\bm e(t))={\rm sgn}(\bm e(t)){\rm sgn}(\bm e(t))^{*}=\|{\rm sgn}(\bm e(t))\|_{1},
	\\
	&(4)\ {\rm sgn}(\bm e(t))^{*}(\bm e(t))^{c}+((\bm e(t))^{c})^{*}{\rm sgn}(\bm e(t))= 2\|(\bm e(t))^{c}\|_{1}\geq \left\{
	\begin{aligned}
		2n^{1-c}\|(\bm e(t))\|_{1}^{c}, & \ c>1,
		\\ 
		2\|(\bm e(t))\|_{1}^{c}, \quad & 0<c\leq 1,\
	\end{aligned}
	\right.
	\\
	&(5)\ \|\int_{t-\pi}^{t}\bm f(\bm e(s))ds\|_{1}\leq \int_{t-\pi}^{t}\|\bm f(\bm e(s))\|_{1}ds.
\end{flalign*}

\section{Main results}
In this section, by designing effective controllers, some sufficient conditions are established to achieve the synchronization of the drive system \eqref{2.1} and the response system \eqref{2.2} in fixed time. The one norm of the quaternion is to demonstrate the practicability of our method by achieving FXTSYN of UCQVMNNs.

To synchronize the drive system and the response system in a fixed time, a novel controller $\bm u_{p}(t)$ is designed as follows
\begin{equation}\label{3.1}
	\bm u_{p}(t)=\lambda_{1p}\bm e_{p}(t)
	-\lambda_{2p}\bm e_{p}(t)^{\alpha}-\lambda_{3p}\bm e_{p}(t)^{\beta}+\lambda_{4p}\bm e_{p}(t-\tau(t))+\lambda_{5p}\int_{t-\pi}^{t}\|\bm e_{p}(s)\|_{1}ds,
\end{equation}
where $\lambda_{1p},\lambda_{2p},\lambda_{3p}, \lambda_{4p}$ and $\lambda_{5p}$ are real constants, and the numbers $\alpha,\beta\in \mathbb{R}$ satisfy $0<\alpha<1$ and $\beta>1$, respectively.

\vspace{3ex}
\noindent \textbf{Theorem 3.1}. Suppose that Assumptions $(A_{1})$ holds. If parameters $\lambda_{1p},\lambda_{2p},\lambda_{3p}, \lambda_{4p}$ and $\lambda_{5p}$ in the controller \eqref{3.1} satisfy
\begin{align}
	&d_{p}-\lambda_{1p}-\sum_{q=1}^{n}\upsilon_{p}a^{+}_{qp}\geq0,\nonumber
	\\
	&\sum_{q=1}^{n}\varrho_{p}b^{+}_{qp}+\lambda_{4p}\leq0,\nonumber
	\\
	&\sum_{q=1}^{n}\iota_{p}c^{+}_{qp}+\lambda_{5p}\leq0,
	\quad \lambda_{2p}>0, \quad \lambda_{3p}>0,
	\label{3.2}
\end{align}
for $p,q=1,2,...,n$, then the drive system \eqref{2.1} synchronizes to the response system \eqref{2.2} in a fixed time. The settling time is estimated as follows
\begin{equation}\label{3.3}
	\begin{aligned}
		T_{1}=
		\frac{1}{\lambda_{2}}
		\Big(\frac{\lambda_{2}}{n^{2(1-\beta)}\lambda_{3}}\Big)
		^{\frac{1-\alpha}{\beta-\alpha}}
		\Big(\frac{1}{\beta-1}+\frac{1}{1-\alpha}\Big).
	\end{aligned}
\end{equation}
where $\lambda_{2}=\min\limits_{1\leq p\leq n}\{\lambda_{2p}\}$, $\lambda_{3}=\min\limits_{1\leq p\leq n}\{\lambda_{3p}\}$.

\vspace{3ex}
\noindent{\bf Proof.} Consider the Lyapunov function
\begin{equation}\label{3.4}
	V(t)=\frac{1}{2}\sum_{p=1}^{n}({\rm sgn}(\bm e_{p}(t))^{*}\bm e_{p}(t)+\bm e_{p}(t)^{*}{\rm sgn}(\bm e_{p}(t))).
\end{equation}
By applying Assumption (A1) to the derivative along the trajectory of the error system \eqref{2.7}, one can obtain
\begin{align*}
	\frac{d}{dt}V(t)
	&=\frac{1}{2}\sum_{p=1}^{n}({\rm sgn}(\bm e_{p}(t))^{*}\frac{d}{dt}\bm e_{p}(t)+(\frac{d}{dt}\bm e_{p}(t))^{*}{\rm sgn}(\bm e_{p}(t)))
	\\
	&=\frac{1}{2}\sum_{p=1}^{n}{\rm sgn}(\bm e_{p}(t))^{*}\Big\{-d_{p}\bm e_{p}(t)+
	\sum_{q=1}^{n}\tilde{\bm a}_{pq}(t)\big[\bm f_{q}(\bm y_{q}(t))-\bm f_{q}(\bm x_{q}(t))\big]
	\\
	& \quad
	+\sum_{q=1}^{n}\tilde{\bm b}_{pq}(t)\big[\bm g_{q}(\bm y_{q}(t-\tau(t)))-\bm g_{q}(\bm x_{q}(t-\tau(t)))\big]
		\\
	& \quad
	+\sum_{q=1}^{n}\tilde{\bm c}_{pq}(t)\int_{t-\pi}^{t}{(\bm h_{q}(\bm y_{q}(s))-\bm h_{q}(\bm x_{q}(s)))}ds
	\\
	& \quad
	+\lambda_{1p}\bm e_{p}(t)
	-\lambda_{2p}(\bm e_{p}(t))^{\alpha}
	-\lambda_{3p}(\bm e_{p}(t))^{\beta}
	+\lambda_{4p}\bm e_{p}(t-\tau(t))
	+\lambda_{5p}\int_{t-\pi}^{t}\|\bm e_{p}(s)\|_{1}ds\Big\}
	\\
	& \quad
	+\frac{1}{2}\sum_{p=1}^{n}\Big\{-d_{p}\bm e_{p}(t)+
	\sum_{q=1}^{n}\tilde{\bm a}_{pq}(t)\big[\bm f_{q}(\bm y_{q}(t))-\bm f_{q}(\bm x_{q}(t))\big]
	\\
	& \quad
	+\sum_{q=1}^{n}\tilde{\bm b}_{pq}(t)\big[\bm g_{q}(\bm y_{q}(t-\tau(t)))-\bm g_{q}(\bm x_{q}(t-\tau(t)))\big]
		\\
		& \quad
	+\sum_{q=1}^{n}\tilde{\bm c}_{pq}(t)\int_{t-\pi}^{t}{(\bm h_{q}(\bm y_{q}(s))-\bm h_{q}(\bm x_{q}(s)))}ds
	\\
	& \quad
	+\lambda_{1p}\bm e_{p}(t)
	-\lambda_{2p}(\bm e_{p}(t))^{\alpha}
	-\lambda_{3p}(\bm e_{p}(t))^{\beta}
	+\lambda_{4p}\bm e_{p}(t-\tau(t))
	\\
	& \quad
	+\lambda_{5p}\int_{t-\pi}^{t}\|\bm e_{p}(s)\|_{1}ds\Big\}^{*}{\rm sgn}(\bm e_{p}(t))
	\\
	&=\frac{1}{2}\sum_{p=1}^{n}\Big[{\rm sgn}(\bm e_{p}(t))^{*}(-d_{p})\bm e_{p}(t)+(\bm e_{p}(t))^{*}(-d_{p})^{*}{\rm sgn}(\bm e_{p}(t))\Big]
	\\
	& \quad
	+\frac{1}{2}\sum_{p=1}^{n}\sum_{q=1}^{n}\Big[{\rm sgn}(\bm e_{p}(t))^{*}\tilde{\bm a}_{pq}(t)[\bm f_{q}(\bm y_{q}(t))-\bm f_{q}(\bm x_{q}(t))]
	\\
	& \quad
	+[\bm f_{q}(\bm y_{q}(t))-\bm f_{q}(\bm x_{q}(t))]^{*}(\tilde{\bm a}_{pq}(t))^{*}{\rm sgn}(\bm e_{p}(t))\Big]
	\\
	& \quad
	+\frac{1}{2}\sum_{p=1}^{n}\sum_{q=1}^{n}\Big[{\rm sgn}(\bm e_{p}(t))^{*}\tilde{\bm b}_{pq}(t)[\bm g_{q}(\bm y_{q}(t-\tau(t)))-\bm g_{q}(\bm x_{q}(t-\tau(t)))]
	\\
	& \quad
	+[\bm g_{q}(\bm y_{q}(t-\tau(t)))-\bm g_{q}(\bm x_{q}(t-\tau(t)))]^{*}(\tilde{\bm b}_{pq}(t))^{*}{\rm sgn}(\bm e_{p}(t))\Big]
	\\
	& \quad
	+\frac{1}{2}\sum_{p=1}^{n}\sum_{q=1}^{n}\Big[{\rm sgn}(\bm e_{p}(t))^{*}\tilde{\bm c}_{pq}(t)\big(\int_{t-\pi}^{t}{(\bm h_{q}(\bm y_{q}(s))-\bm h_{q}(\bm x_{q}(s)))}ds\big)
	\\
	& \quad
	+\big(\int_{t-\pi}^{t}{(\bm h_{q}(\bm y_{q}(s))-\bm h_{q}(\bm x_{q}(s)))}ds\big)^{*}(\tilde{\bm c}_{pq}(t))^{*}{\rm sgn}(\bm e_{p}(t))\Big]
	\\
	& \quad
	+\frac{1}{2}\sum_{p=1}^{n}\lambda_{1p}\Big[{\rm sgn}(\bm e_{p}(t))^{*}\bm e_{p}(t)+\bm e_{p}(t)^{*}{\rm sgn}(\bm e_{p}(t))\Big]
	\\
	& \quad
	-\frac{1}{2}\sum_{p=1}^{n}\lambda_{2p}\Big[{\rm sgn}(\bm e_{p}(t))^{*}(\bm e_{p}(t))^{\alpha}+((\bm e_{p}(t))^{\alpha})^{*}{\rm sgn}(\bm e_{p}(t))\Big]
	\\
	& \quad
	-\frac{1}{2}\sum_{p=1}^{n}\lambda_{3p}\Big[{\rm sgn}(\bm e_{p}(t))^{*}(\bm e_{p}(t))^{\beta}+((\bm e_{p}(t))^{\beta})^{*}{\rm sgn}(\bm e_{p}(t))\Big]
	\\
	& \quad
	+\frac{1}{2}\sum_{p=1}^{n}\lambda_{4p}\Big[{\rm sgn}(\bm e_{p}(t))^{*}\bm e_{p}(t-\tau(t))+(\bm e_{q}(t-\tau(t)))^{*}{\rm sgn}(\bm e_{p}(t))\Big]
	\\
	& \quad
	+\frac{1}{2}\sum_{p=1}^{n}\lambda_{5p}\Big[{\rm sgn}(\bm e_{p}(t))^{*}\big(\int_{t-\pi}^{t}\|\bm e_{p}(s)\|_{1}ds\big)
	+\big(\int_{t-\pi}^{t}\|\bm e_{p}(s)\|_{1}ds\big)^{*}{\rm sgn}(\bm e_{p}(t))\Big]
	\\
	&\leq
	\frac{1}{2}\sum_{p=1}^{n}(-d_{p}+\lambda_{1p})\Big[{\rm sgn}(\bm e_{p}(t))^{*}\bm e_{p}(t)+(\bm e_{p}(t))^{*}{\rm sgn}(\bm e_{p}(t))\Big]
	\\
	& \quad
	+\frac{1}{2}\sum_{p=1}^{n}\sum_{q=1}^{n}\Big[{\rm sgn}(\bm e_{p}(t))^{*}\tilde{\bm a}_{pq}(t)\upsilon_{q}\|\bm e_{q}(t)\|_{1}+[\upsilon_{q}\|\bm e_{q}(t)\|_{1}]^{*}(\tilde{\bm a}_{pq}(t))^{*}{\rm sgn}(\bm e_{p}(t))\Big]
	\\
	& \quad
	+\frac{1}{2}\sum_{p=1}^{n}\sum_{q=1}^{n}\Big[{\rm sgn}(\bm e_{p}(t))^{*}\tilde{\bm b}_{pq}(t)\varrho_{q}\|\bm e_{q}(t-\tau(t))\|_{1}
	\\
	& \quad
	+[\varrho_{q}\|\bm e_{q}(t-\tau(t))\|_{1}]^{*}(\tilde{\bm b}_{pq}(t))^{*}{\rm sgn}(\bm e_{p}(t))\Big]
	\\
	& \quad
	+\frac{1}{2}\sum_{p=1}^{n}\sum_{q=1}^{n}\Big[{\rm sgn}(\bm e_{p}(t))^{*}\tilde{\bm c}_{pq}(t)\big(\int_{t-\pi}^{t}(\iota_{q}\|\bm e_{q}(s)\|_{1})ds\big)
	\\
	& \quad
	+\big(\int_{t-\pi}^{t}(\iota_{q}\|\bm e_{q}(s)\|_{1})ds\big)^{*}(\tilde{\bm c}_{pq}(t))^{*}{\rm sgn}(\bm e_{p}(t))\Big]
	\\
	& \quad
	-\frac{1}{2}\sum_{p=1}^{n}\lambda_{2p}\Big[{\rm sgn}(\bm e_{p}(t))^{*}(\bm e_{p}(t))^{\alpha}+((\bm e_{p}(t))^{\alpha})^{*}{\rm sgn}(\bm e_{p}(t))\Big]
	\\
	& \quad
	-\frac{1}{2}\sum_{p=1}^{n}\lambda_{3p}\Big[{\rm sgn}(\bm e_{p}(t))^{*}(\bm e_{p}(t))^{\beta}+((\bm e_{p}(t))^{\beta})^{*}{\rm sgn}(\bm e_{p}(t))\Big]
	\\
	& \quad
	+\frac{1}{2}\sum_{p=1}^{n}\lambda_{4p}\Big[{\rm sgn}(\bm e_{p}(t))^{*}\bm e_{p}(t-\tau(t))+(\bm e_{p}(t-\tau(t)))^{*}{\rm sgn}(\bm e_{p}(t))\Big]
	\\
	& \quad
	+\frac{1}{2}\sum_{p=1}^{n}\lambda_{5p}\Big[{\rm sgn}(\bm e_{p}(t))^{*}\big(\int_{t-\pi}^{t}\|\bm e_{p}(s)\|_{1}ds\big)
	+\big(\int_{t-\pi}^{t}\|\bm e_{p}(s)\|_{1}ds\big)^{*}{\rm sgn}(\bm e_{p}(t))\Big]
\end{align*}
By Lemma 2.4 and Assumption (A1), we have
\begin{flalign}\label{3.5}
	&\frac{1}{2}\sum_{p=1}^{n}(-d_{p}+\lambda_{1p})\Big[{\rm sgn}(\bm e_{p}(t))^{*}\bm e_{p}(t)+(\bm e_{p}(t))^{*}{\rm sgn}(\bm e_{p}(t))\Big]
	=-\sum_{p=1}^{n}(d_{p}-\lambda_{1p})\|\bm e_{p}(t)\|_{1},
	\\
	&
	\frac{1}{2}\sum_{p=1}^{n}\sum_{q=1}^{n}\Big[{\rm sgn}(\bm e_{p}(t))^{*}\tilde{\bm a}_{pq}(t)\upsilon_{q}\|\bm e_{q}(t)\|_{1}+[\upsilon_{q}\|\bm e_{q}(t)\|_{1}]^{*}(\tilde{\bm a}_{pq}(t))^{*}{\rm sgn}(\bm e_{p}(t))\Big]\nonumber
	\\
	&\leq
	\sum_{p=1}^{n}\sum_{q=1}^{n}\upsilon_{q}a^{+}_{pq}
	\|\bm e_{q}(t)\|_{1}
	=\sum_{p=1}^{n}\sum_{q=1}^{n}\upsilon_{p}a^{+}_{qp}
	\|\bm e_{p}(t)\|_{1},
\end{flalign}
\begin{flalign}\label{Q1}
	&\frac{1}{2}\sum_{p=1}^{n}\sum_{q=1}^{n}\Big[{\rm sgn}(\bm e_{p}(t))^{*}\tilde{\bm b}_{pq}(t)\varrho_{q}\|\bm e_{q}(t-\tau(t))\|_{1}
	+[\varrho_{q}\|\bm e_{q}(t-\tau(t))\|_{1}]^{*}(\tilde{\bm b}_{pq}(t))^{*}{\rm sgn}(\bm e_{p}(t))\Big]\nonumber
	\\
	& 
	+\frac{1}{2}\sum_{p=1}^{n}\lambda_{4p}\Big[{\rm sgn}(\bm e_{p}(t))^{*}\bm e_{p}(t-\tau(t))+(\bm e_{p}(t-\tau(t)))^{*}{\rm sgn}(\bm e_{p}(t))\Big]\nonumber
	\\
	& \leq
	\sum_{p=1}^{n}\sum_{q=1}^{n}\varrho_{q}b^{+}_{pq}\|\bm e_{q}(t-\tau(t))\|_{1}+\sum_{p=1}^{n}\lambda_{4p}\|\bm e_{p}(t-\tau(t))\|_{1}\nonumber
	\\
	&=\sum_{p=1}^{n}\big(\sum_{q=1}^{n}\varrho_{p}b^{+}_{qp}+\lambda_{4p}\big)\|\bm e_{p}(t-\tau(t))\|_{1}.
\end{flalign}
And
\begin{flalign}\label{3.6}
	& \frac{1}{2}\sum_{p=1}^{n}\sum_{q=1}^{n}\Big[{\rm sgn}(\bm e_{p}(t))^{*}\tilde{\bm c}_{pq}(t)\big(
	\int_{t-\pi}^{t}(\iota_{q}\|\bm e_{q}(s)\|_{1})ds\big)\nonumber
	\\
	&
	+\big(\int_{t-\pi}^{t}(\iota_{q}\|\bm e_{q}(s)\|_{1})ds\big)^{*}(\tilde{\bm c}_{pq}(t))^{*}{\rm sgn}(\bm e_{p}(t))\Big]\nonumber
	\\ 
	&+\frac{1}{2}\sum_{p=1}^{n}\lambda_{5p}\Big[{\rm sgn}(\bm e_{p}(t))^{*}\big(\int_{t-\pi}^{t}\|\bm e_{p}(s)\|_{1}ds\big)
	+\big(\int_{t-\pi}^{t}\|\bm e_{p}(s)\|_{1}ds\big)^{*}{\rm sgn}(\bm e_{p}(t))\Big]\nonumber
	\\
	& \leq
	\sum_{p=1}^{n}\sum_{q=1}^{n}\iota_{q}c^{+}_{pq}
	\int_{t-\pi}^{t}\|\bm e_{q}(s)\|_{1}ds+\sum_{p=1}^{n}\lambda_{5p}\int_{t-\pi}^{t}\|\bm e_{p}(s)\|_{1}ds\nonumber
	\\
	&=\sum_{p=1}^{n}(\sum_{q=1}^{n}\iota_{p}c^{+}_{qp}+\lambda_{5p})\int_{t-\pi}^{t}\|\bm e_{p}(s)\|_{1}ds.
\end{flalign}
Analogously, using Lemma 2.3 and Lemma 2.4, it is not difficult to get the following conclusion
\begin{flalign}\label{3.7}
	& 
	-\frac{1}{2}\sum_{p=1}^{n}\lambda_{2p}\Big[{\rm sgn}(\bm e_{p}(t))^{*}(\bm e_{p}(t))^{\alpha}+((\bm e_{p}(t))^{\alpha})^{*}{\rm sgn}(\bm e_{p}(t))\Big]
	\leq
	-\sum_{p=1}^{n}\lambda_{2p}\|\bm e_{p}(t)\|_{1}^{\alpha}, \nonumber
	\\
	& 
	-\frac{1}{2}\sum_{p=1}^{n}\lambda_{3p}\Big[{\rm sgn}(\bm e_{p}(t))^{*}(\bm e_{p}(t))^{\beta}+((\bm e_{p}(t))^{\beta})^{*}{\rm sgn}(\bm e_{p}(t))\Big]
	\leq
	-n^{1-\beta}\sum_{p=1}^{n}\lambda_{3p}\|\bm e_{p}(t)\|_{1}^{\beta}. %
\end{flalign}

Combined with the above formulas \eqref{3.5}-\eqref{3.7} and Lemma 2.3, one can
obtain
\begin{align*}
	\frac{d}{dt}V(t)&
	\leq-\sum_{p=1}^{n}\big(d_{p}-\lambda_{1p}-\sum_{q=1}^{n}\upsilon_{p}a^{+}_{qp}\big)\|\bm e_{p}(t)\|_{1}
	+\sum_{p=1}^{n}\big(\sum_{q=1}^{n}\varrho_{p}b^{+}_{qp}+\lambda_{4p}\big)\|\bm e_{p}(t-\tau(t))\|_{1}\nonumber
	\\
	& \quad
	+\sum_{p=1}^{n}(\sum_{q=1}^{n}\iota_{p}c^{+}_{qp}+\lambda_{5p})\int_{t-\pi}^{t}\|\bm e_{p}(s)\|_{1}ds
	-\sum_{p=1}^{n}\lambda_{2p}\|\bm e_{p}(t)\|_{1}^{\alpha}
	\\
	& \quad
	-n^{1-\beta}\sum_{p=1}^{n}\lambda_{3p}\|\bm e_{p}(t)\|_{1}^{\beta}\nonumber
	\\
	& 
	\leq -\sum_{p=1}^{n}\lambda_{2p}\|\bm e_{p}(t)\|_{1}^{\alpha}
	-n^{1-\beta}\sum_{p=1}^{n}\lambda_{3p}\|\bm e_{p}(t)\|_{1}^{\beta}\nonumber
	\\
	& 
	\leq -\lambda_{2}\sum_{p=1}^{n}\|\bm e_{p}(t)\|_{1}^{\alpha}
	-n^{1-\beta}\lambda_{3}\sum_{p=1}^{n}\|\bm e_{p}(t)\|_{1}^{\beta}\nonumber
	\\
	& 
	\leq -\lambda_{2}(\sum_{p=1}^{n}\|\bm e_{p}(t)\|_{1})^{\alpha}
	-n^{2(1-\beta)}\lambda_{3}(\sum_{p=1}^{n}\|\bm e_{p}(t)\|_{1})^{\beta}\nonumber
	\\
	& 
	= -\lambda_{2}(V(t))^{\alpha}
	-n^{2(1-\beta)}\lambda_{3}(V(t))^{\beta},
\end{align*}
where $\lambda_{2}=\min\limits_{1\leq p\leq n}\{\lambda_{2p}\}$, $\lambda_{3}=\min\limits_{1\leq p\leq n}\{\lambda_{3p}\}$.

Based on Lemma 2.1, it follows that the error system \eqref{2.7}  is fixed-time stable, that is the drive system \eqref{2.1} and the response system \eqref{2.2} achieve synchronization in fixed-time with the controller \eqref{3.1}.
Furthermore, one can estimate the settling time by the following equality
\begin{equation}\label{3.10}
	\begin{aligned}
		T_{1}=
		\frac{1}{\lambda_{2}}
		\Big(\frac{\lambda_{2}}{n^{2(1-\beta)}\lambda_{3}}\Big)
		^{\frac{1-\alpha}{\beta-\alpha}}
		\Big(\frac{1}{\beta-1}+\frac{1}{1-\alpha}\Big).
	\end{aligned}
\end{equation}
The proof is completed. \qquad \qquad \qquad \qquad \qquad \qquad \qquad \qquad \qquad \qquad \qquad \qquad \quad$\Box$

\vspace{1ex}
\noindent \textbf{Corollary 3.1.}
By Remark 2.1, taking the same conditions and the controller as in Theorem 3.1, one can obtain that system \eqref{2.1} and system \eqref{2.2} are synchronized within a fixed time. Furthermore, the settling time can be estimated as
\begin{equation}\label{3.11}
	\begin{aligned}
		T_{2}=
		\frac{1}{\lambda_{2}(1-\alpha)}
		+\frac{1}{n^{2(1-\beta)}\lambda_{3}(\beta-1)}.
	\end{aligned}
\end{equation}

\vspace{1ex}

It is well known that the error system can be stabilized within a fixed time, that is, $e(t)\rightarrow 0$ within time T, by developing an appropriate controller and Lyapunov function. In general, it can be seen from the inequality 
$\frac{d}{dt}V(e(t)) \leq {-aV^{\alpha}(e(t))}-{bV^{\beta}(e(t))}$ with $0<\alpha<1, \beta>1$ in Lemma 2.1 that the right-hand side contains two terms: the index of one term is bigger than $0$ and smaller than $1$, while the index of the other one is larger than 1. In the controller \eqref{3.1}, we can see that the first, fourth, and fifth items are designed to allow the error system to achieve Lyapunov stability. The second and third terms are designed to achieve synchronization of the drive-response system in a fixed time. And then compute the settling time using the parameters of these two items. However, in Lemma 2.1 and Remark 2.1, the settling time estimate includes both errors from greater than $1$ to $1$ and then from $1$ to $0$. This must be considered because the two terms exist in the inequality at the same time, whether they play a role or not, which may cause the estimated settling time to be inaccurate.

Li et al. \cite{li2020fixed} proposed a novel Lemma 2.2 to guarantee fixed-time synchronization of discontinuous neural networks, where $V(t)\geq 1$ or $0<V(t)<1$ is first judged and the functioning part is intelligently chosen to be more economical. It is clear that Lemma 2.2 is an excellent alternative to existing techniques and can significantly reduce energy consumption while achieving a more precise settling time than most relevant research. For the first time, Lemma 2.2 is used to synchronize the UCQVMNNs in a fixed time.

According to Lemma 2.2, in order to make the error system \eqref{2.7} stable in fixed time, different from Theorem 3.1, a new nonlinear controller is designed as follows:
\begin{equation}\label{3.13}
	\bm u_{p}(t)=-k_{1p}\bm e_{p}(t)+k_{2p}\bm e_{p}(t-\tau(t))
	-\mu (\bm e_{p}(t))^{\gamma+{\rm sgn}(\|\bm e(t)\|_{1}-1)}
	+k_{3p}\sum_{q=1}^{n}\int_{t-\pi}^{t}\|\bm e_{q}(s)\|_{1}ds,
\end{equation}
where $\bm e(t)=(\bm e_{1}(t), \bm e_{2}(t),\cdots, \bm e_{n}(t))^{T}\in \mathbb{H}^{n}$, parameter $\gamma\in \mathbb{R}$ satisfies $1\leq\gamma<2$, the feedback gains $k_{1p},k_{2p}$ and $\mu$ are real constants, and $k_{1p} \geq 0$, $\mu>0$ will be determined later.
\vspace{2ex}

\noindent \textbf{Theorem 3.2.} Suppose that Assumption ($A1$) holds, and  for
constants $k_{2P},$ $k_{3P}$ and $\mu$, the following inequalities are fulfilled 
\begin{align}\label{3.14}
	&d<\mu_{1},\nonumber 
	\\
	&\sum_{q=1}^{n}\varrho_{p}b^{+}_{qp}+k_{2p}\leq 0,\nonumber
	\\
	&\sum_{q=1}^{n}\iota_{q}c^{+}_{pq}+k_{3p}\leq 0,
\end{align}
then, UCQVMNNs \eqref{2.1} and \eqref{2.2} can achieve FIXSYN with controller \eqref{3.13}. Furthermore, for any initial condition, the settling time can be estimated by
\begin{align}
	&T_{4}=\left\{
	\begin{aligned}
		\frac{1}{d(2-\gamma)}\ln\frac{\mu}{\mu-d}+\frac{1}{\gamma(\mu_{1}-d)},\quad\ d>0,
		\\
		\frac{1}{\mu(2-\gamma)}+\frac{1}{\mu_{1}\gamma},\ \qquad \qquad \qquad \quad d=0,
		\\
		\quad	\frac{1}{d(2-\gamma)}\ln\frac{\mu}{\mu-d}+\frac{1}{d\gamma}\ln\frac{\mu_{1}}{\mu_{1}-d},\quad d<0,
	\end{aligned}
	\right.
	\label{3.15}
\end{align} 

\noindent{ In particular}, if the initial value is less than $1$, the corresponding settling time is estimated as follows
\begin{align}
	&T_{3}=\left\{
	\begin{aligned}
		\frac{1}{d(2-\gamma)}\ln\frac{\mu}{\mu-d}+\frac{1}{\gamma(\mu-d)},\quad\ d>0,
		\\
		\frac{1}{\mu(2-\gamma)}+\frac{1}{\mu\gamma},\ \qquad \qquad \qquad \quad d=0,
		\\
		\quad	\frac{1}{d(2-\gamma)}\ln\frac{\mu}{\mu-d}+\frac{1}{d\gamma}\ln\frac{\mu}{\mu-d},\quad d<0,
	\end{aligned}
	\right.
	\label{3.16}
\end{align} 
where $d=\max\limits_{1\leq p,q\leq n}\{-d_{p}-k_{1p}+\sum_{q=1}^{n}\upsilon_{p}a^{+}_{qp}\}$, $\mu_{1}=\mu n^{-2\gamma}$.

\noindent{\bf Proof.} Consider the Lyapunov function
\begin{equation}\label{3.17}
	V(t)=\frac{1}{2}\sum_{p=1}^{n}({\rm sgn}(\bm e_{p}(t))^{*}\bm e_{p}(t)+\bm e_{p}(t)^{*}{\rm sgn}(\bm e_{p}(t))),\\
\end{equation}
According to Assumption (A1), calculating the derivative along the trajectory of the error system \eqref{2.7}, one can get
\begin{align*}
	\frac{d}{dt}V(t)
	&=\frac{1}{2}\sum_{p=1}^{n}({\rm sgn}(\bm e_{p}(t))^{*}\frac{d}{dt}\bm e_{p}(t)+(\frac{d}{dt}\bm e_{p}(t))^{*}{\rm sgn}(\bm e_{p}(t)))
	\\
	&=\frac{1}{2}\sum_{p=1}^{n}{\rm sgn}(\bm e_{p}(t))^{*}\Big\{-d_{p}\bm e_{p}(t)
	+
	\sum_{q=1}^{n}\tilde{\bm a}_{pq}(t)\big[\bm f_{q}(\bm y_{q}(t))-\bm f_{q}(\bm x_{q}(t))\big]
	\\
	& \quad
	+\sum_{q=1}^{n}\tilde{\bm b}_{pq}(t)\big[\bm g_{q}(\bm y_{q}(t-\tau(t)))-\bm g_{q}(\bm x_{q}(t-\tau(t)))\big]
	\\
	& \quad
	+\sum_{q=1}^{n}\tilde{\bm c}_{pq}(t)\int_{t-\pi}^{t}{(\bm h_{q}(\bm y_{q}(s))-\bm h_{q}(\bm x_{q}(s)))}ds
	\\
	& \quad
	-k_{1p}\bm e_{p}(t)
	+k_{2p}\bm e_{p}(t-\tau(t))
	-\mu(\bm e_{p}(t))^{\gamma+{\rm sgn}(\|\bm e_{p}(t)\|_{1}-1)}
	+k_{3p}\sum_{q=1}^{n}\int_{t-\pi}^{t}\|\bm e_{q}(s)\|_{1}ds\Big\}
	\\
	& \quad
	+\frac{1}{2}\sum_{p=1}^{n}\Big\{-d_{p}\bm e_{p}(t)
	+
	\sum_{q=1}^{n}\tilde{\bm a}_{pq}(t)\big[\bm f_{q}(\bm y_{q}(t))-\bm f_{q}(\bm x_{q}(t))\big]
	\\
	& \quad
	+\sum_{q=1}^{n}\tilde{\bm b}_{pq}(t)\big[\bm g_{q}(\bm y_{q}(t-\tau(t)))-\bm g_{q}(\bm x_{q}(t-\tau(t)))\big]
	\\
	& \quad
	+\sum_{q=1}^{n}\tilde{\bm c}_{pq}(t)\int_{t-\pi}^{t}{(\bm h_{q}(\bm y_{q}(s))-\bm h_{q}(\bm x_{q}(s)))}ds
	\\
	& \quad
	-k_{1p}\bm e_{p}(t)
	+k_{2p}\bm e_{p}(t-\tau(t))
	-\mu(\bm e_{p}(t))^{\gamma+{\rm sgn}(\|\bm e_{p}(t)\|_{1}-1)}
	\\
	& \quad
	+k_{3p}\sum_{q=1}^{n}\int_{t-\pi}^{t}\|\bm e_{q}(s)\|_{1}ds\Big\}^{*}{\rm sgn}(\bm e_{p}(t))
	\\
	&=
	\frac{1}{2}\sum_{p=1}^{n}\Big[{\rm sgn}(\bm e_{p}(t))^{*}(-d_{p})\bm e_{p}(t)+(\bm e_{p}(t))^{*}(-d_{p})^{*}{\rm sgn}(\bm e_{p}(t))\Big]
	\\
	& \quad
	+\frac{1}{2}\sum_{p=1}^{n}\sum_{q=1}^{n}\Big[{\rm sgn}(\bm e_{p}(t))^{*}\tilde{\bm a}_{pq}(t)[\bm f_{q}(\bm y_{q}(t))-\bm f_{q}(\bm x_{q}(t))]
	\\
	& \quad
	+[\bm f_{q}(\bm y_{q}(t))-\bm f_{q}(\bm x_{q}(t))]^{*}(\tilde{\bm a}_{pq}(t))^{*}{\rm sgn}(\bm e_{p}(t))\Big]
	\\
	& \quad
	+\frac{1}{2}\sum_{p=1}^{n}\sum_{q=1}^{n}\Big[{\rm sgn}(\bm e_{p}(t))^{*}\tilde{\bm b}_{pq}(t)[\bm g_{q}(\bm y_{q}(t-\tau(t)))-\bm g_{q}(\bm x_{q}(t-\tau(t)))]
	\\
	& \quad
	+[\bm g_{q}(\bm y_{q}(t-\tau(t)))-\bm g_{q}(\bm x_{q}(t-\tau(t)))]^{*}(\tilde{\bm b}_{pq}(t))^{*}{\rm sgn}(\bm e_{p}(t))\Big]
	\\
	& \quad
	+\frac{1}{2}\sum_{p=1}^{n}\sum_{q=1}^{n}\Big[{\rm sgn}(\bm e_{p}(t))^{*}\tilde{\bm c}_{pq}(t)\big(\int_{t-\pi}^{t}{(\bm h_{q}(\bm y_{q}(s))-\bm h_{q}(\bm x_{q}(s)))}ds\big)
	\\
	& \quad
	+\big(\int_{t-\pi}^{t}{(\bm h_{q}(\bm y_{q}(s))-\bm h_{q}(\bm x_{q}(s)))}ds\big)^{*}(\tilde{\bm c}_{pq}(t))^{*}{\rm sgn}(\bm e_{p}(t))\Big]
	\\
	& \quad
	-\frac{1}{2}\sum_{p=1}^{n}k_{1p}\Big[{\rm sgn}(\bm e_{p}(t))^{*}\bm e_{p}(t)+\bm e_{p}(t)^{*}{\rm sgn}(\bm e_{p}(t))\Big]
	\\
	& \quad
	-\frac{1}{2}\sum_{p=1}^{n}\mu\Big[{\rm sgn}(\bm e_{p}(t))^{*}(\bm e_{p}(t))^{\gamma+{\rm sgn}(\|\bm e(t)\|_{1}-1)}+((\bm e_{p}(t))^{\gamma+{\rm sgn}(\|\bm e(t)\|_{1}-1)})^{*}{\rm sgn}(\bm e_{p}(t))\Big]
	\\
	& \quad
	+\frac{1}{2}\sum_{p=1}^{n}k_{2p}\Big[{\rm sgn}(\bm e_{p}(t))^{*}\bm e_{p}(t-\tau(t))+(\bm e_{p}(t-\tau(t)))^{*}{\rm sgn}(\bm e_{p}(t))\Big]
	\\
	& \quad
	+\frac{1}{2}\sum_{p=1}^{n}\sum_{q=1}^{n}k_{3p}\Big[{\rm sgn}(\bm e_{p}(t))^{*}\big(\int_{t-\pi}^{t}\|\bm e_{q}(s)\|_{1}ds\big)
	+\big(\int_{t-\pi}^{t}\|\bm e_{q}(s)\|_{1}ds\big)^{*}){\rm sgn}(\bm e_{p}(t))\Big]
	\\
	&\leq
	\frac{1}{2}\sum_{p=1}^{n}\Big\{(-d_{p}-k_{1p})\Big[{\rm sgn}(\bm e_{p}(t))^{*}\bm e_{p}(t)+(\bm e_{p}(t))^{*}{\rm sgn}(\bm e_{p}(t))\Big]
	\Big\}
	\\
	& \quad
	+\frac{1}{2}\sum_{p=1}^{n}\sum_{q=1}^{n}\Big[{\rm sgn}(\bm e_{p}(t))^{*}\tilde{\bm a}_{pq}(t)\upsilon_{q}\|\bm e_{q}(t)\|_{1}+[\upsilon_{q}\|\bm e_{q}(t)\|_{1}]^{*}(\tilde{\bm a}_{pq}(t))^{*}{\rm sgn}(\bm e_{p}(t))\Big]
	\\
	& \quad
	+\frac{1}{2}\sum_{p=1}^{n}\sum_{q=1}^{n}\Big[{\rm sgn}(\bm e_{p}(t))^{*}\tilde{\bm b}_{pq}(t)\varrho_{q}\|\bm e_{q}(t-\tau(t))\|_{1}
	\\
	& \quad
	+[\varrho_{q}\|\bm e_{q}(t-\tau(t))\|_{1}]^{*}(\tilde{\bm b}_{pq}(t))^{*}{\rm sgn}(\bm e_{p}(t))\Big]
	\\
	& \quad
	+\frac{1}{2}\sum_{p=1}^{n}k_{2p}\Big[{\rm sgn}(\bm e_{p}(t))^{*}\bm e_{p}(t-\tau(t))+(\bm e_{p}(t-\tau(t)))^{*}{\rm sgn}(\bm e_{p}(t))\Big]
	\\
	& \quad
	+\frac{1}{2}\sum_{p=1}^{n}\sum_{q=1}^{n}\Big[{\rm sgn}(\bm e_{p}(t))^{*}\tilde{\bm c}_{pq}(t)\big(\int_{t-\pi}^{t}(\iota_{q}\|\bm e_{q}(s)\|_{1})ds\big)
	\\
	& \quad
	+\big(\int_{t-\pi}^{t}(\iota_{q}\|\bm e_{q}(s)\|_{1})ds\big)^{*}(\tilde{\bm c}_{pq}(t))^{*}{\rm sgn}(\bm e_{p}(t))\Big]
	\\
	& \quad
	+\frac{1}{2}\sum_{p=1}^{n}\sum_{q=1}^{n}k_{3p}\Big[{\rm sgn}(\bm e_{p}(t))^{*}\big(\int_{t-\pi}^{t}\|\bm e_{q}(s)\|_{1}ds\big)
	+\big(\int_{t-\pi}^{t}\|\bm e_{q}(s)\|_{1}ds\big)^{*}){\rm sgn}(\bm e_{p}(t))\Big]
	\\
	& \quad
	-\frac{1}{2}\mu\sum_{p=1}^{n}\Big[{\rm sgn}(\bm e_{p}(t))^{*}(\bm e_{p}(t))^{\gamma+{\rm sgn}(\|\bm e(t)\|_{1}-1)}+((\bm e_{p}(t))^{\gamma+{\rm sgn}(\|\bm e(t)\|_{1}-1)})^{*}{\rm sgn}(\bm e_{p}(t))\Big].
\end{align*}

Using Lemma 2.4 can be get
\begin{align}\label{3.18}
	& \quad 
	\frac{1}{2}\sum_{p=1}^{n}\sum_{q=1}^{n}\Big[{\rm sgn}(\bm e_{p}(t))^{*}\tilde{\bm c}_{pq}(t)\big(\int_{t-\pi}^{t}(\iota_{q}\|\bm e_{q}(s)\|_{1})ds\big) \nonumber
	\\
	& \quad
	+\big(\int_{t-\pi}^{t}(\iota_{q}\|\bm e_{q}(s)\|_{1})ds\big)^{*}(\tilde{\bm c}_{pq}(t))^{*}{\rm sgn}(\bm e_{p}(t))\Big]\nonumber
	\\
	& \quad
	+\frac{1}{2}\sum_{p=1}^{n}\sum_{q=1}^{n}k_{3p}\Big[{\rm sgn}(\bm e_{p}(t))^{*}\big(\int_{t-\pi}^{t}\|\bm e_{q}(s)\|_{1}ds\big)
	+\big(\int_{t-\pi}^{t}\|\bm e_{q}(s)\|_{1}ds\big)^{*}){\rm sgn}(\bm e_{p}(t))\Big]\nonumber
	\\
	& 
	\leq \sum_{p=1}^{n}\sum_{q=1}^{n}(\iota_{q}c^{+}_{pq}+k_{3p})\int_{t-\pi}^{t}\|\bm e_{q}(s)\|_{1}ds.
\end{align}

Next, in combination with Lemma 2.2, we will discuss two cases according to the error value.

Case 1: when $0<\|\bm e(t)\|_{1}<1$ ($0<V(t)<1$), so that ${\rm sgn}(\|\bm e(t)\|_{1}-1)=-1$, $0\leq\gamma+{\rm sgn}(\|\bm e(t)\|_{1}-1)<1$. According to the conclusion in Theorem 1 and Lemma 2.3 can be directly obtained following inequalities.
\begin{align}\label{3.19}
	& \quad 
	-\frac{1}{2}\Big[{\rm sgn}(\bm e_{p}(t))^{*}(\bm e_{p}(t))^{\gamma+{\rm sgn}(\|\bm e(t)\|_{1}-1)}+((\bm e_{p}(t))^{\gamma+{\rm sgn}(\|\bm e(t)\|_{1}-1)})^{*}{\rm sgn}(\bm e_{p}(t))\Big]\nonumber
	\\
	& 
	\leq 
	-\|\bm e_{p}(t)\|_{1}^{\gamma+{\rm sgn}(\|\bm e(t)\|_{1}-1)}.
\end{align}

Therefore combined with the above formulas \eqref{3.5}-\eqref{Q1}, \eqref{3.18} and \eqref{3.19}, then using Lemma 2.3 and Lemma 2.4, one can obtain
\begin{align}\label{3.20}
	\frac{d}{dt}V(t)&
	\leq\sum_{p=1}^{n}\big(-d_{p}-k_{1p}+\sum_{q=1}^{n}\upsilon_{p}a^{+}_{qp}\big)\|\bm e_{p}(t)\|_{1}
	+\sum_{p=1}^{n}\big(\sum_{q=1}^{n}\varrho_{p}b^{+}_{qp}+k_{2p}\big)\|\bm e_{p}(t-\tau(t))\|_{1}\nonumber
	\\
	& \quad
	+\sum_{p=1}^{n}\sum_{q=1}^{n}(\iota_{q}c^{+}_{pq}+k_{3p})\int_{t-\pi}^{t}\|\bm e_{q}(s)\|_{1}ds
	-\mu\sum_{p=1}^{n}\|\bm e_{p}(t)\|_{1}^{\gamma+{\rm sgn}(\|\bm e(t)\|_{1}-1)}\nonumber
	\\
	& 
	\leq d\sum_{p=1}^{n}\|\bm e_{p}(t)\|_{1}
	-\mu\sum_{p=1}^{n}\|\bm e_{p}(t)\|_{1}^{\gamma+{\rm sgn}(\|\bm e(t)\|_{1}-1)}\nonumber
	\\
	& 
	\leq
	d\sum_{p=1}^{n}\|\bm e_{p}(t)\|_{1}
	-\mu(\sum_{p=1}^{n}\|\bm e_{p}(t)\|_{1})^{\gamma+{\rm sgn}(\|\bm e(t)\|_{1}-1)}\nonumber
	\\
	& 
	= dV(t)-\mu(V(t))^{\gamma+{\rm sgn}(V(t)-1)},
\end{align}
where $d=\max\limits_{1\leq p,q\leq n}\{-d_{p}-k_{1p}+\sum_{q=1}^{n}\upsilon_{p}a^{+}_{qp}\}$.

With conditions in \eqref{3.14} holding, based on Lemma 2.2, it follows that the error system \eqref{2.7} is fixed-time stable, that is the QVMNNs \eqref{2.1} and the QVMNNs \eqref{2.2} achieve synchronization in fixed-time with the controller \eqref{3.13}. And the settling time is estimated as
\begin{align}\nonumber
	&T_{3}=\left\{
	\begin{aligned}
		\frac{1}{d(2-\gamma)}\ln\frac{\mu}{\mu-d}+\frac{1}{\gamma(\mu-d)},\quad\ d>0,
		\\
		\frac{1}{\mu(2-\gamma)}+\frac{1}{\mu\gamma},\ \qquad \qquad \qquad \quad d=0,
		\\
		\quad	\frac{1}{d(2-\gamma)}\ln\frac{\mu}{\mu-d}+\frac{1}{d\gamma}\ln\frac{\mu}{\mu-d},\quad d<0,
	\end{aligned}
	\right.
\end{align} 

Case 2: when $\|\bm e(t)\|_{1}\geq 1$ ($V(t)\geq1$), so that ${\rm sgn}(\|\bm e(t)\|_{1}-1)\geq 0$, $\gamma+{\rm sgn}(\|\bm e(t)\|_{1}-1)\geq \gamma \geq1$. According to Lemma 2.3 and Lemma 2.4, it can ba seen that
\begin{align}\label{3.21}
	& \quad 
	-\frac{1}{2}\Big[{\rm sgn}(\bm e_{p}(t))^{*}(\bm e_{p}(t))^{\gamma+{\rm sgn}(\|\bm e(t)\|_{1}-1)}+((\bm e_{p}(t))^{\gamma+{\rm sgn}(\|\bm e(t)\|_{1}-1)})^{*}{\rm sgn}(\bm e_{p}(t))\Big]\nonumber
	\\
	& 
	\leq 
	-n^{1-\gamma-{\rm sgn}(\|\bm e(t)\|_{1}-1)}\|\bm e_{p}(t)\|_{1}^{\gamma+{\rm sgn}(\|\bm e(t)\|_{1}-1)}\nonumber
	\\
	& 
	\leq -n^{-\gamma}\|\bm e_{p}(t)\|_{1}^{\gamma+{\rm sgn}(\|\bm e(t)\|_{1}-1)}
\end{align}
Therefore we have
\begin{align}\nonumber
	\frac{d}{dt}V(t)&
	\leq\sum_{p=1}^{n}\big(-d_{p}-k_{1p}+\sum_{q=1}^{n}\upsilon_{p}a^{+}_{qp}\big)\|\bm e_{p}(t)\|_{1}
	+\sum_{p=1}^{n}\big(\sum_{q=1}^{n}\varrho_{p}b^{+}_{qp}+k_{2p}\big)\|\bm e_{p}(t-\tau(t))\|_{1}\nonumber
	\\
	& \quad
	+\sum_{p=1}^{n}\sum_{q=1}^{n}(\iota_{p}c^{+}_{qp}+k_{3p})\int_{t-\pi}^{t}\|\bm e_{q}(s)\|_{1}ds
	-\mu\sum_{p=1}^{n}n^{-\gamma}\|\bm e_{p}(t)\|_{1}^{\gamma+{\rm sgn}(\|\bm e(t)\|_{1}-1)}\nonumber
	\\
	& 
	\leq d\sum_{p=1}^{n}\|\bm e_{p}(t)\|_{1}
	-\mu\sum_{p=1}^{n}n^{-\gamma}\|\bm e_{p}(t)\|_{1}^{\gamma+{\rm sgn}(\|\bm e(t)\|_{1}-1)}\nonumber
	\\
	& 
	\leq
	d\sum_{p=1}^{n}\|\bm e_{p}(t)\|_{1}
	-\mu n^{-\gamma}n^{1-\gamma-{\rm sgn}(\|\bm e(t)\|_{1}-1)}(\sum_{p=1}^{n}\|\bm e_{p}(t)\|_{1})^{\gamma+{\rm sgn}(\|\bm e(t)\|_{1}-1)}\nonumber
	\\
	& 
	\leq
	d\sum_{p=1}^{n}\|\bm e_{p}(t)\|_{1}
	-\mu n^{-2\gamma}(\sum_{p=1}^{n}\|\bm e_{p}(t)\|_{1})^{\gamma+{\rm sgn}(\|\bm e(t)\|_{1}-1)}\nonumber
	\\
	& 
	= dV(t)-\mu_{1}(V(t))^{\gamma+{\rm sgn}(V(t)-1)},
\end{align}
where $\mu_{1}=\mu n^{-2\gamma}$. In order to guarantee the Lyapunov stability, an extra condition $d<\mu_{1}$ must be satisfied.

Analogously, by Lemma 2.2, we can get that the UCQVMNNs \eqref{2.1} and \eqref{2.2} achieve synchronization in fixed-time with the controller \eqref{3.13}.
And the settling time can be inferred as
\begin{align}\nonumber
	&T_{4}=\left\{
	\begin{aligned}
		\frac{1}{d(2-\gamma)}\ln\frac{\mu}{\mu-d}+\frac{1}{\gamma(\mu_{1}-d)},\quad\ d>0,
		\\
		\frac{1}{\mu(2-\gamma)}+\frac{1}{\mu_{1}\gamma},\ \qquad \qquad \qquad \quad d=0,
		\\
		\quad	\frac{1}{d(2-\gamma)}\ln\frac{\mu}{\mu-d}+\frac{1}{d\gamma}\ln\frac{\mu_{1}}{\mu_{1}-d},\quad d<0,
	\end{aligned}
	\right.
\end{align} 
The proof is completed. \qquad \qquad \qquad \qquad \qquad \qquad \qquad \qquad \qquad \qquad \qquad \qquad \quad $\Box$ 

{ \begin{remark}
		Different from Theorem 3.1, Theorem 3.2 does not require parameter $d<0$. Different $d$ values yield different settling time estimates, resulting in greater selectivity and robustness. If we take a proper $k_{1p} > 0$ to make $d < 0$, any value of positive $\mu$ can guarantee the existence of condition \eqref{3.14}. If $k_{1p} = 0$, for $ p=1,2,...,n$ in controller \eqref{3.13}, maybe $d > 0$, so we should choose a large enough positive $\mu$ to make inequality \eqref{3.14} hold. In this case, controller \eqref{3.13} only has three terms. As a result, it is critical to choose an appropriate $k_{1p}$ for the given situation.
	\end{remark}
}


\section{Numerical Simulations}
In this section, three numerical instances are given to illustrate the effectiveness of our theoretical results obtained in the previous section.

\vspace{3ex}
\noindent \textbf{Example 1}. Consider the following 2-dimensional UCQVMNNs with mixed delays as the drive system
\begin{align}
	&\frac{d}{dt}\bm x_{p}(t)=
	-d_{p}\bm x_{p}(t)+\sum_{q=1}^{2}\bm  a_{pq}(\bm x_{p}(t))f_{q}(\bm x_{q}(t))
	+\sum_{q=1}^{2} \bm b_{pq}(\bm x_{p}(t))g_{q}(\bm x_{q}(t-\tau(t)))\nonumber
	\\
	&\qquad\quad 
	+\sum_{q=1}^{2}\bm c_{pq}(\bm x_{p}(t))\int_{t-\pi}^{t}{\bm h_{q}(\bm x_{q}(s))}ds,
	\label{4.1}
\end{align}
the response system is
\begin{align}
	&\frac{d}{dt}\bm y_{p}(t)=
	-d_{p}\bm y_{p}(t)+\sum_{q=1}^{2} a_{pq}(\bm y_{p}(t))\bm f_{q}(\bm y_{q}(t))
	+\sum_{q=1}^{2}\bm b_{pq}(\bm y_{p}(t))\bm g_{q}(\bm y_{q}(t-\tau(t)))\nonumber
	\\
	&\qquad\quad \
	+\sum_{q=1}^{2}\bm c_{pq}(\bm y_{p}(t))\int_{t-\pi}^{t}{\bm h_{q}(\bm y_{q}(s))}ds+\bm u_{p}(t),
	\label{4.2}
\end{align}
where $p,q=1,2$, $d_{p}=0.5,$ 
$\tau(t)=0.3{\rm sin}(t)+0.4$, $\pi=0.4$, so let $\tau=0.7$, and the memristive connection weights are

\begin{equation} \nonumber       
	\bm A=\left(                 
	\begin{array}{ccc}   		
		1.8-1.6i-2.3j-1k &  -1-1.5i-1.7j+1.3k\\  		
		1.5+3.5i-2j-1.5k &  -1.5+2i-1.5j-1.6k\\ 		
	\end{array}	
	\right)               
\end{equation}
\begin{equation} \nonumber      
	\bm B=\left(                 
	\begin{array}{ccc}   		
		-0.45+0.3i-0.2j-0.35k & -0.25+0.25i+0.1j-0.4k\\  		
		0.2-0.5i+0.35j-0.25k & 0.2-0.4i-0.18j+0.22k\\ 		
	\end{array}	
	\right)               
\end{equation}
\begin{equation} \nonumber      
	\bm C=\left(                 
	\begin{array}{ccc}   		
		2.4+3i+1j+2k & -1.5+1.2i-2j+1.1k\\  		
		-1+2.1i-1.9j+1.3k & 2+2.3i+1j-3.2k\\ 		
	\end{array}	
	\right)               
\end{equation}
where $|\bm x_{p}(t)|\leq 1$, $p=1,2$.
\begin{equation} \nonumber       
	\bm A=\left(                 
	\begin{array}{ccc}   		
		1.8-1.6i-3.5j-2.3k    &  -0.6+1.5i-1.7j+1.3k\\  		
		-1.9-1.5i-1.7j+1.3k  &  1.5-2i+1.5j+1.2k\\ 		
	\end{array}	
	\right)               
\end{equation}
\begin{equation} \nonumber      
	\bm B=\left(                 
	\begin{array}{ccc}   		
		-0.5+0.3i-0.2j-0.3k & -0.15-0.2i-0.2j+0.45k\\  		
		0.3-0.65i-0.15j+0.1k & -0.44+0.1i+0.16j+0.3k\\ 		
	\end{array}	
	\right)              
\end{equation}
\begin{equation} \nonumber      
	\bm C=\left(                 
	\begin{array}{ccc}   		
		-2.4-2i-1j+1.6k & 1.1-2i+1.5j-1.6k\\  		
		1.4-1.7i+2j-1.4k & -2-2.3i+1j-1.8k\\ 		
	\end{array}	
	\right)               
\end{equation}
where $|\bm x_{p}(t)|> 1$, $p=1,2$.  One can easily compute $a_{11}^{+}=9.2, a_{12}^{+}=5.5, a_{21}^{+}=8.5, a_{22}^{+}=6.6$, $b_{11}^{+}=1.3, b_{12}^{+}=1.0, b_{21}^{+}=1.3, b_{22}^{+}=1.0$ and $c_{11}^{+}=8.4, c_{12}^{+}=6.2, c_{21}^{+}=6.5, c_{22}^{+}=7.1$.
And take the activation functions are
$$\bm f_{q}(\bm x_{q}(t))=2\tanh(x_{q(0)}(t))+2\tanh(x_{q(1)}(t))i+2\tanh(x_{q(2)}(t))j+2\tanh(x_{q(3)}(t))k,$$
$$\bm g_{q}(\bm x_{q}(t))=0.1\tanh(x_{q(0)}(t))+0.1\tanh(x_{q(1)}(t))i+0.1\tanh(x_{q(2)}(t))j+0.1\tanh(x_{q(3)}(t))k,$$
$$\bm h_{q}(\bm x_{q}(t))=0.7\tanh(x_{q(0)}(t))+0.7\tanh(x_{q(1)}(t))i+0.7\tanh(x_{q(2)}(t))j+0.7\tanh(x_{q(3)}(t))k.$$
According to Assumption $(A1)$, a simple calculation yields that  {$\upsilon_{1}=\upsilon_{2}=2, \varrho_{1}=\varrho_{2}=0.1, \iota_{1}=\iota_{2}=0.7.$}

The initial conditions of \eqref{4.1} and \eqref{4.2} are selected as

$\bm\phi_{1}(s)=1.5+2i-0.6j+0.8k,\ \bm \phi_{2}(s)=-1.2-1.5i+1j-0.5k,\ \bm \psi_{1}(s)=2.5-2i-1j+1.2k,\ \bm \psi_{2}(s)=-3+1.6i+0.8j-2k,\ s\in[-0.7,0]$.

If there is no controller, i.e., $\bm u_{1}(t)=\bm u_{2}(t)=0$, the trajectories of the error system is simulated in Fig. 1, which implies that systems \eqref{4.1} and \eqref{4.2} cannot obtain synchronization without control input.

Correspondingly, by \eqref{3.1}, the controllers are designed as follows
\begin{equation}\label{4.3}
	\bm u_{1}(t)=\lambda_{11}\bm e_{1}(t)
	-\lambda_{21}\bm e_{1}^{\alpha}(t)-\lambda_{31}\bm e_{1}^{\beta}(t)+\lambda_{41}\bm e_{1}(t-\tau(t))+\lambda_{51}\int_{t-\pi}^{t}\|\bm e_{1}(s)\|_{1}ds,
\end{equation}
\begin{equation}\label{4.4}
	\bm u_{2}(t)=\lambda_{12}\bm e_{2}(t)-\lambda_{22}\bm e_{2}^{\alpha}(t)-\lambda_{32}\bm e_{2}^{\beta}(t)+\lambda_{42}\bm e_{2}(t-\tau(t))+\lambda_{52}\int_{t-\pi}^{t}\|\bm e_{2}(s)\|_{1}ds,
\end{equation}
\vspace{-0.1in}
\begin{figure}[H]
	\begin{minipage}[t]{0.49\linewidth}
		\centering
		\includegraphics[height=4.5cm, width=8cm]{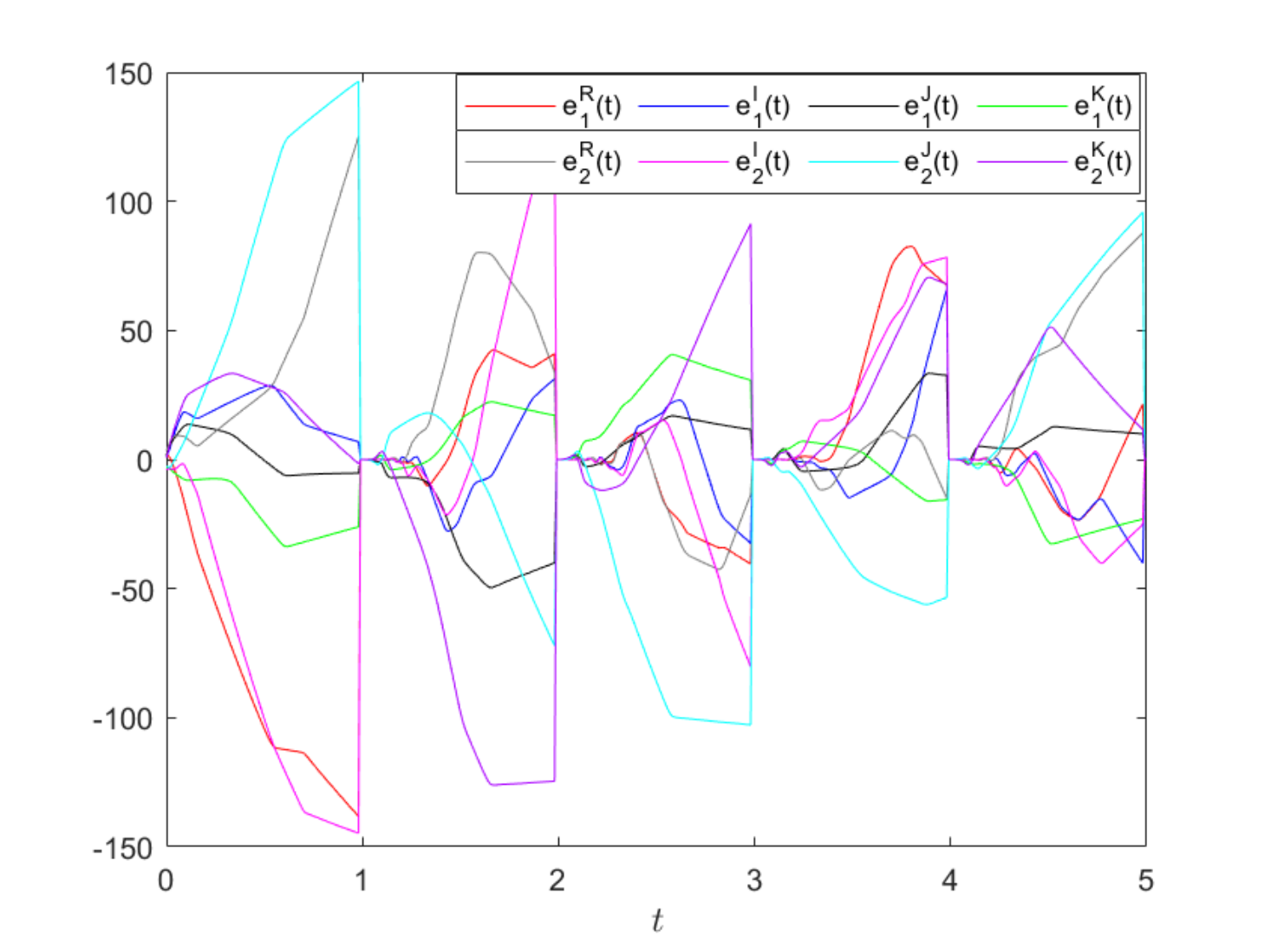}
		\caption{The trajectories of error system without controller.}
	\end{minipage}\quad\hfill
	\begin{minipage}[t]{0.49\linewidth}
		\centering
		\includegraphics[height=4.5cm, width=8cm]{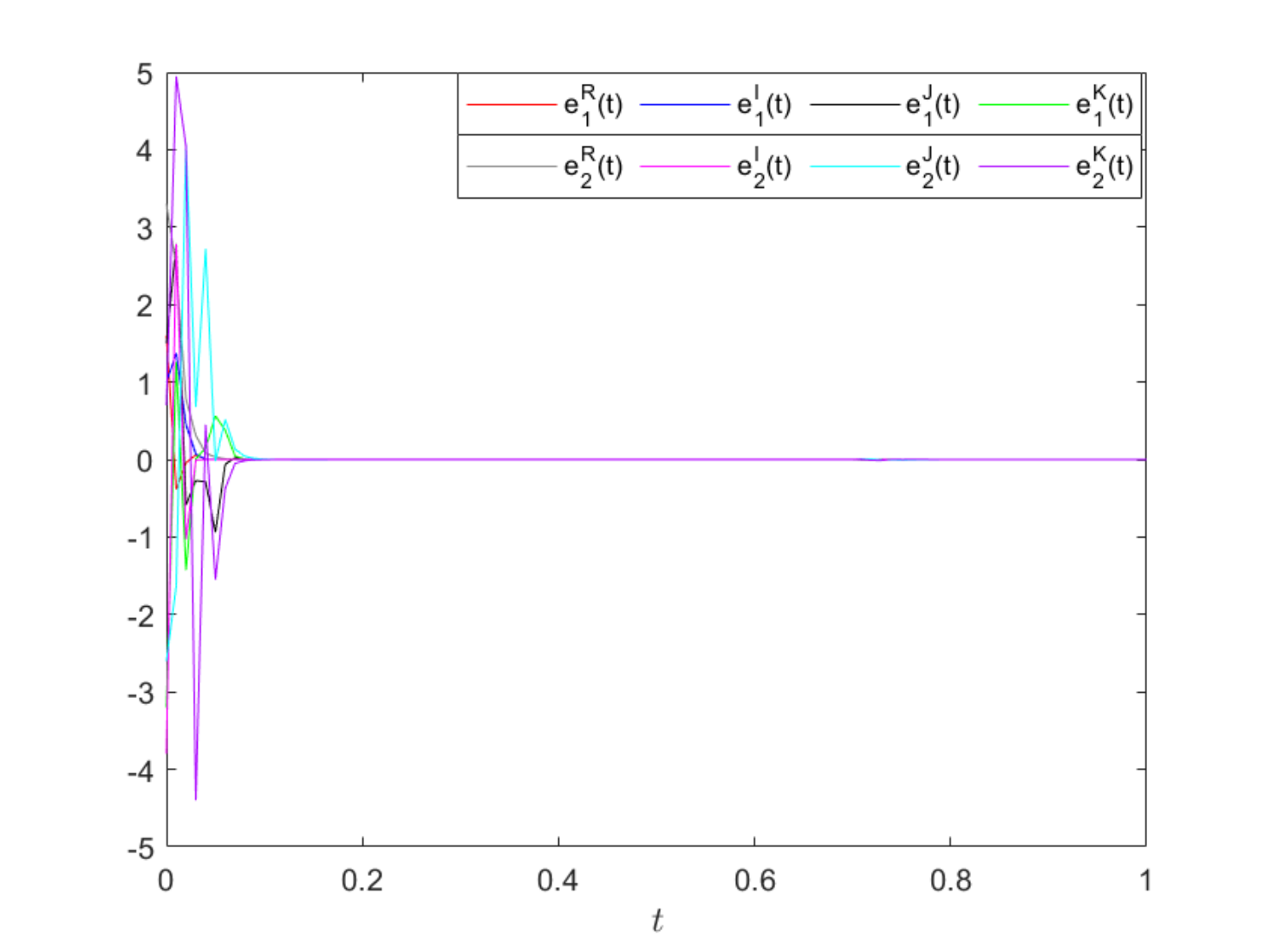}
		\caption{The trajectories of error system with controllers (4.3) and (4.4).}
	\end{minipage}
\end{figure}

The values of the coefficients 
$\lambda_{11}=-80,  \lambda_{12}=-50, \lambda_{21}=\lambda_{22}=1, \lambda_{31}=30, \lambda_{32}=35, \lambda_{41}=-0.26, \lambda_{42}=-0.2,\lambda_{51}=-10.45, \lambda_{52}=-9.35$ can be calculated from \eqref{3.2}, $\alpha=0.6, \beta=1.6$, the conditions in Theorem 3.1 hold. 

\ignore{
\begin{figure}[H]
	\begin{minipage}[t]{0.49\linewidth}
		\centering
		\includegraphics[height=4.5cm, width=8cm]{x1y1-w.PNG}
		\caption{The drive-response system trajectories of $p=1$ without controller.}
	\end{minipage}
	\hfill
	\begin{minipage}[t]{0.49\linewidth}
		\centering
		\includegraphics[height=4.5cm, width=8cm]{x2y2-w.PNG}
		\caption{The drive-response system trajectories of $p=2$ without controller.}
	\end{minipage}
\end{figure}
}

\begin{figure}[H]
	\centering
	\vspace{-0.15in}
	\begin{minipage}{1\linewidth}	
		\subfloat[]{
			\includegraphics[width=0.49\linewidth,height=1.2in]{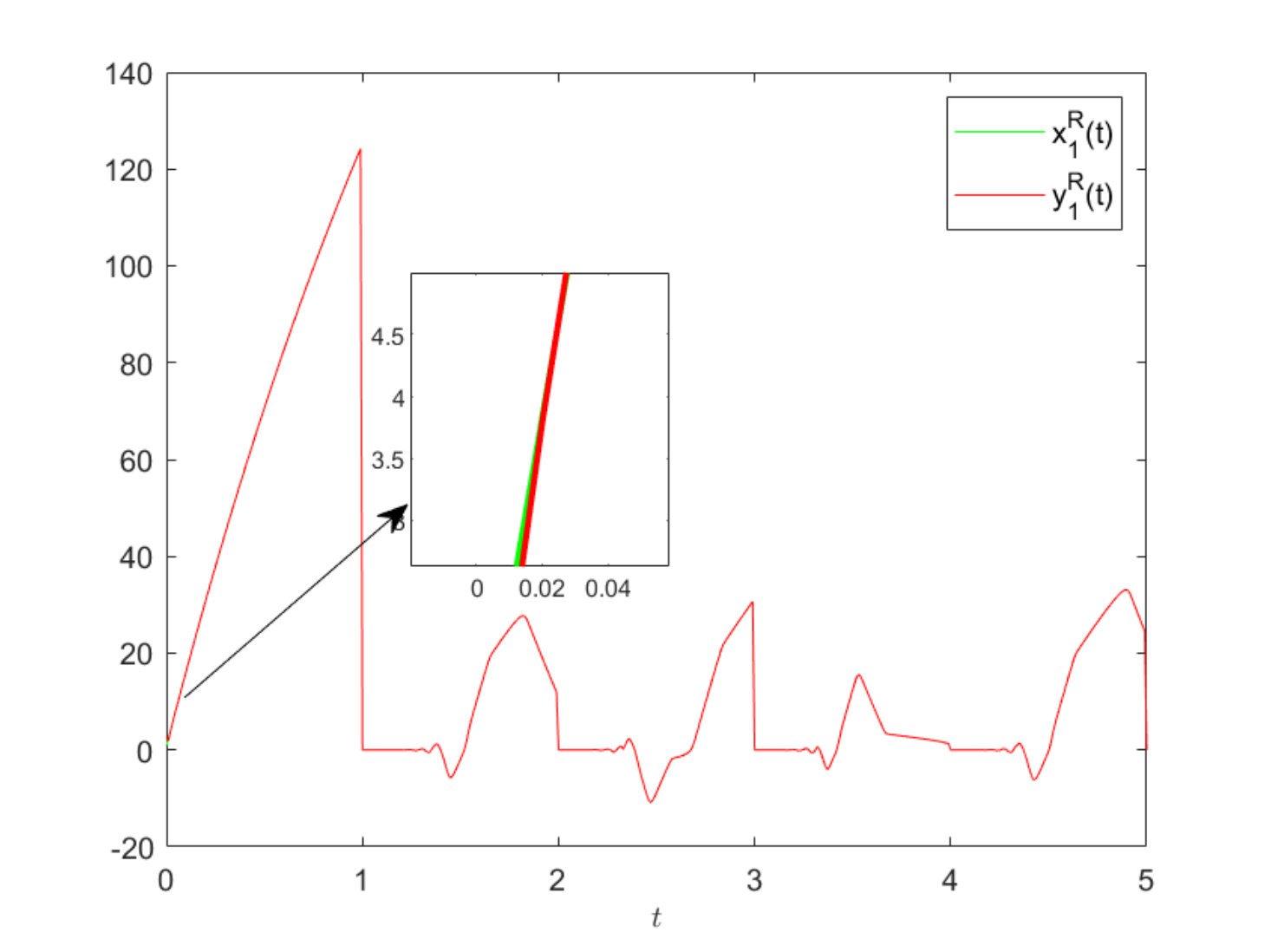}	
		}\noindent
		\subfloat[]{
			\includegraphics[width=0.49\linewidth,height=1.2in]{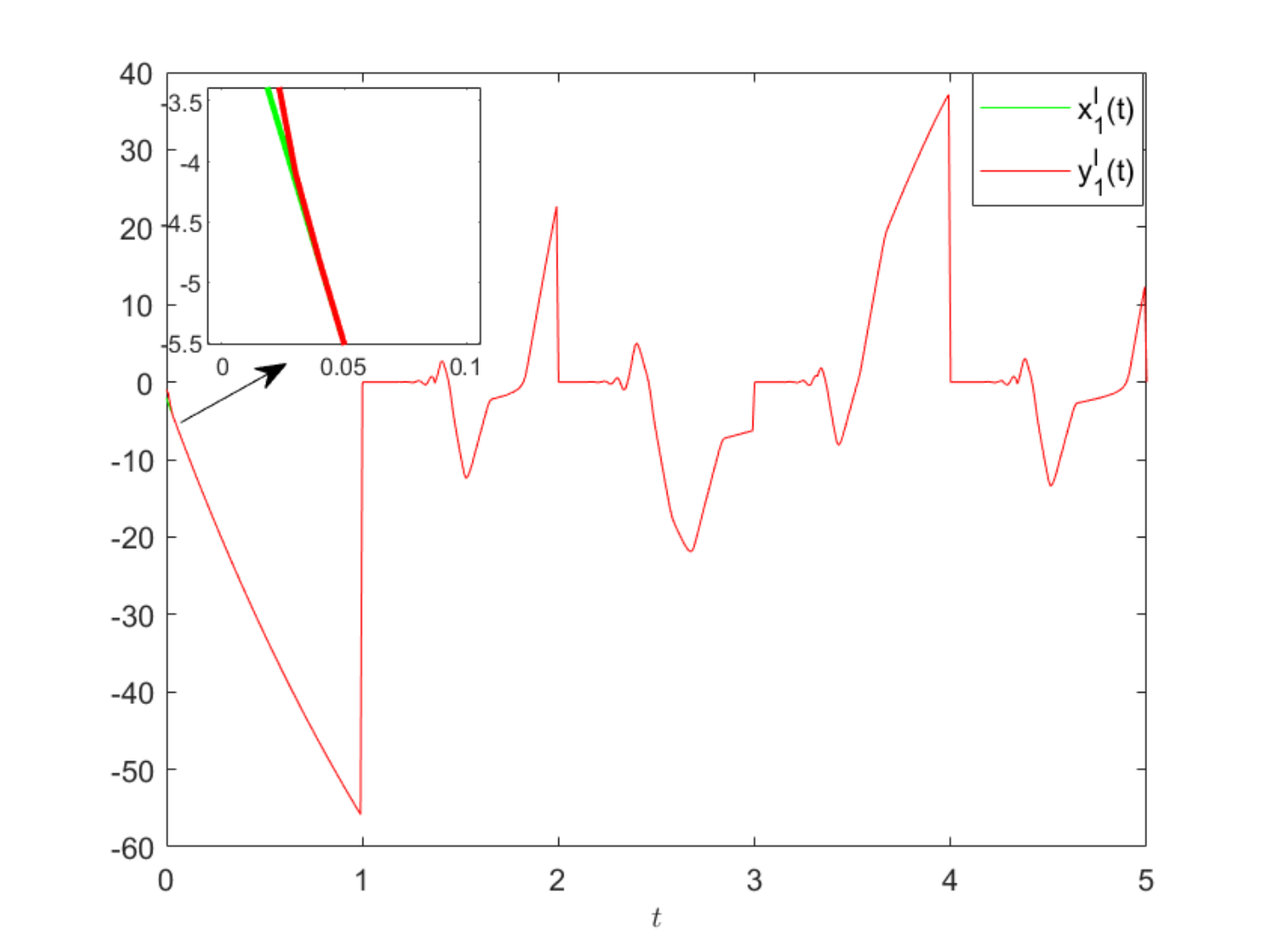}
		}
	\end{minipage}
	\vskip -0.3cm  
	\begin{minipage}{1\linewidth }
		\subfloat[]{
			\includegraphics[width=0.49\linewidth,height=1.2in]{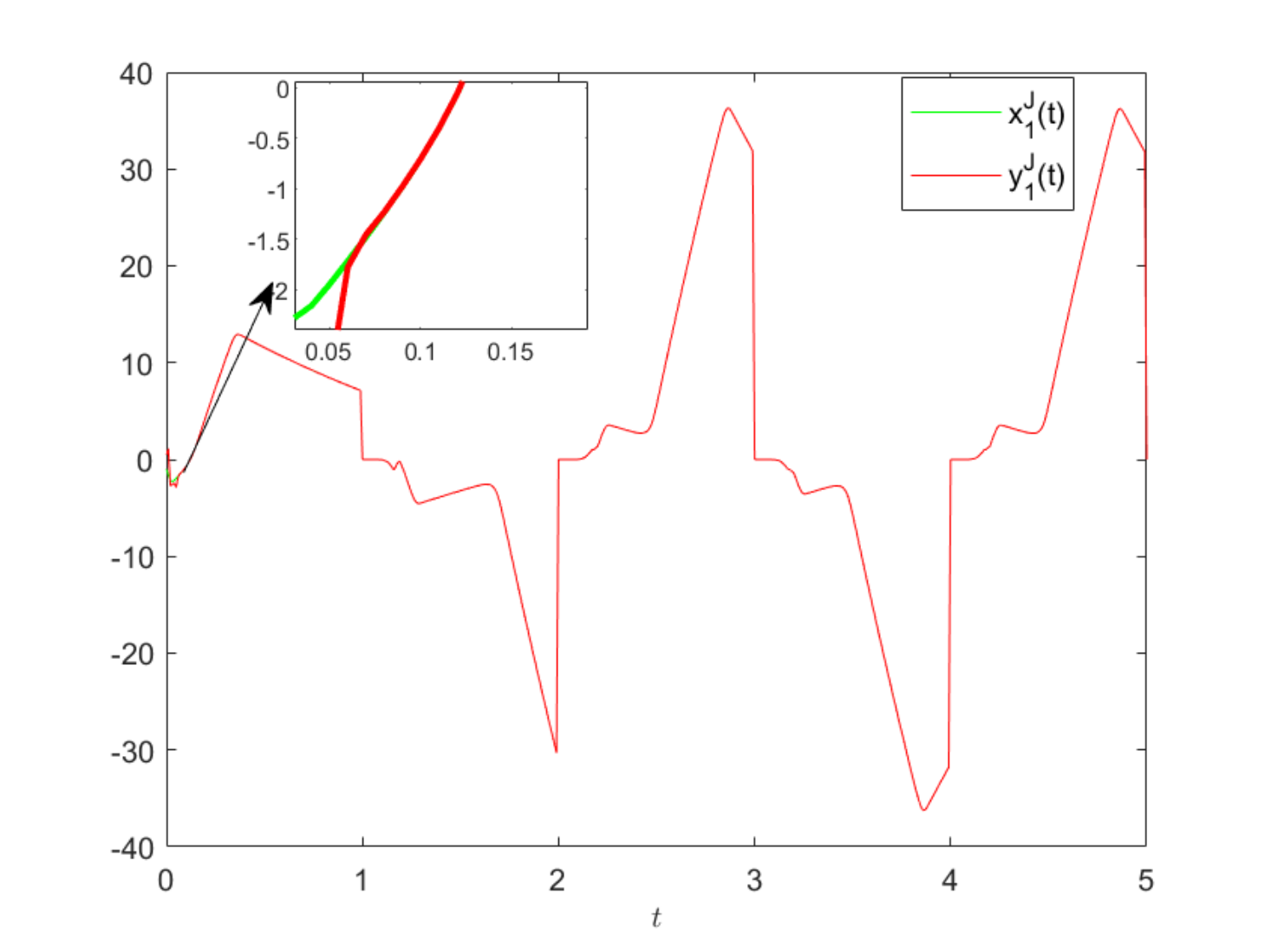}
	}\noindent
	\subfloat[]{
		\includegraphics[width=0.49\linewidth,height=1.2in]{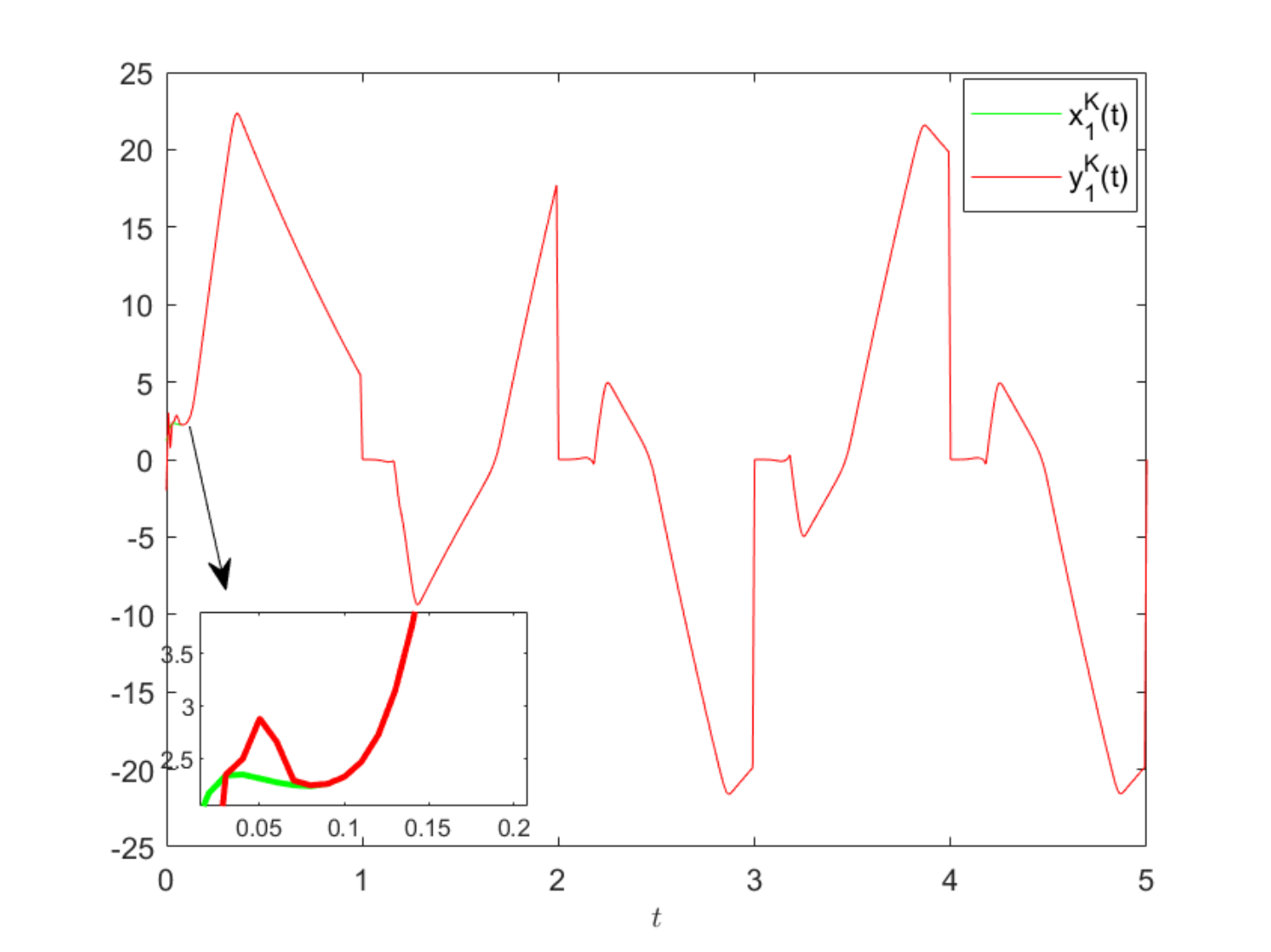}
	}
\end{minipage}
\vspace{-0.12in}
\caption{Trajectories of the real and imaginary parts of drive-response system with controllers (4.3) and (4.4) when $p=1$.}
\vspace{0.0in}	
\end{figure}

Under the designed controllers \eqref{4.3} and \eqref{4.4}, the trajectories of real and imaginary parts of system \eqref{4.1} and \eqref{4.2} are given in Fig. 3-4. These figures indicate that once appending the controllers \eqref{4.3} and \eqref{4.4} to the system \eqref{4.2}, it will be synchronized with the system \eqref{4.1} in a fixed time.

\begin{figure}[H]
\centering
\vspace{-0.15in}
\begin{minipage}{1\linewidth}	
\subfloat[]{
	\includegraphics[width=0.49\linewidth,height=1.2in]{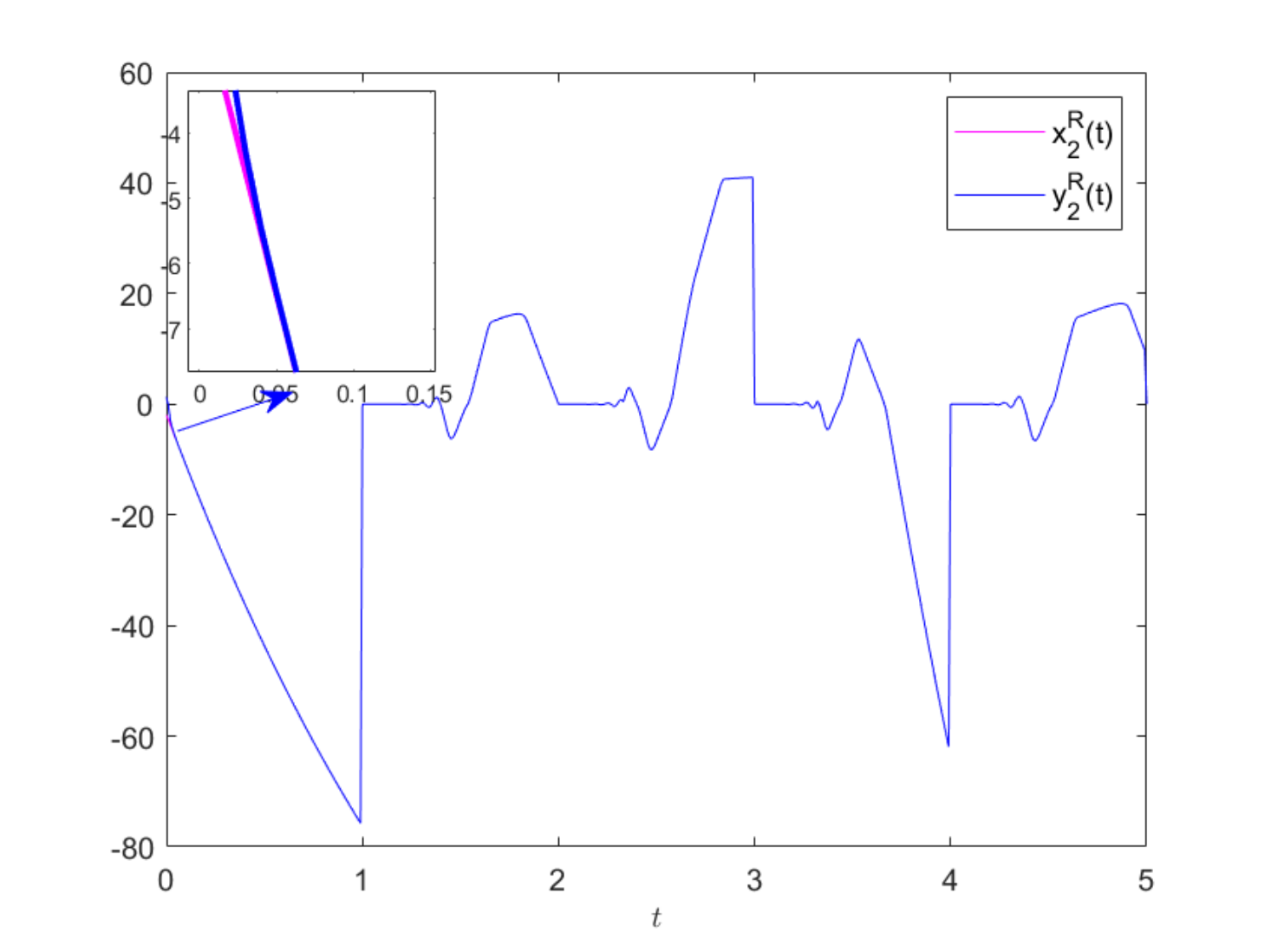}	
}\noindent
\subfloat[]{
	\includegraphics[width=0.49\linewidth,height=1.2in]{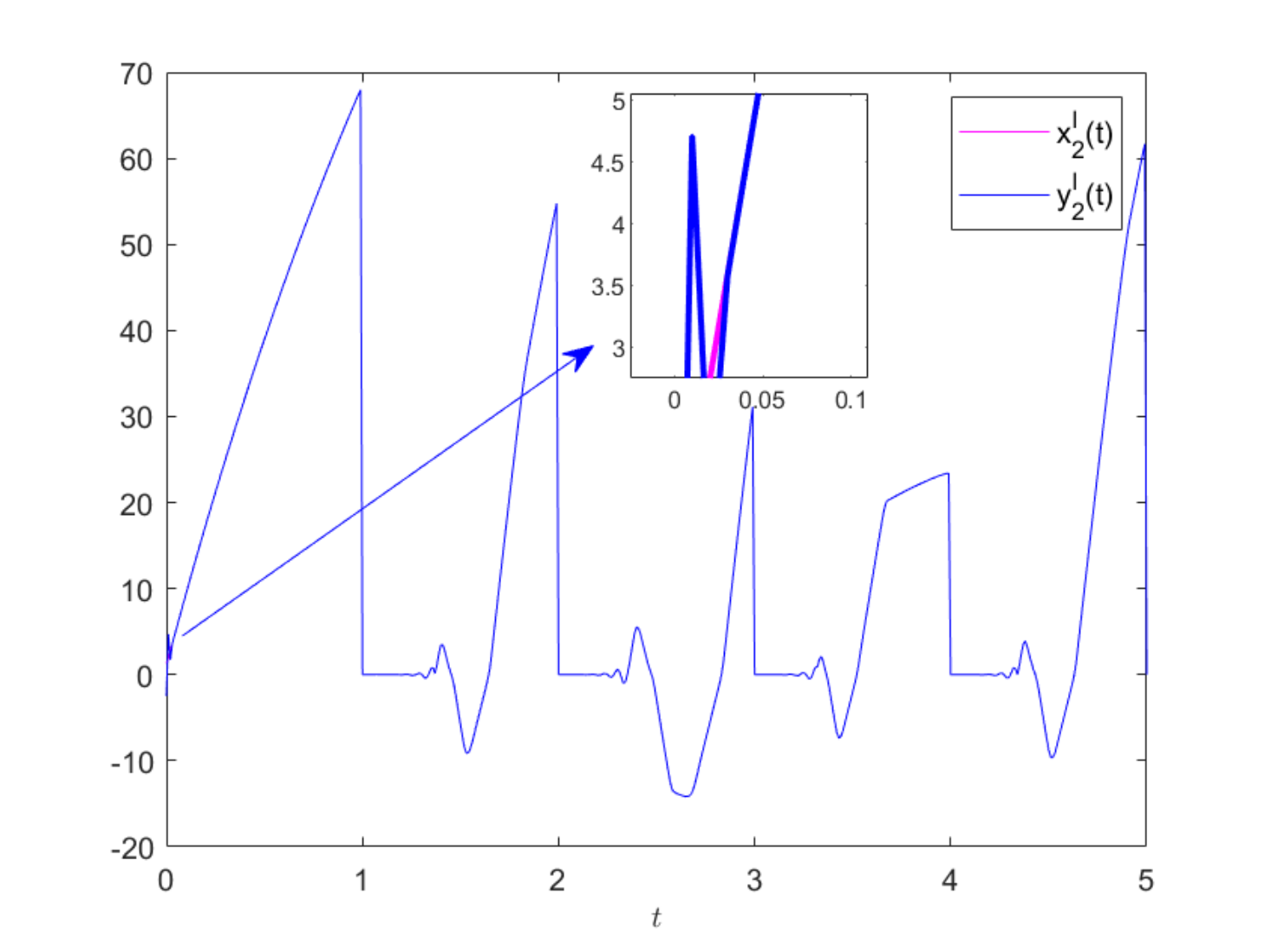}
}
\end{minipage}
\vskip -0.3cm 
\begin{minipage}{1\linewidth }
\subfloat[]{
	\includegraphics[width=0.49\linewidth,height=1.2in]{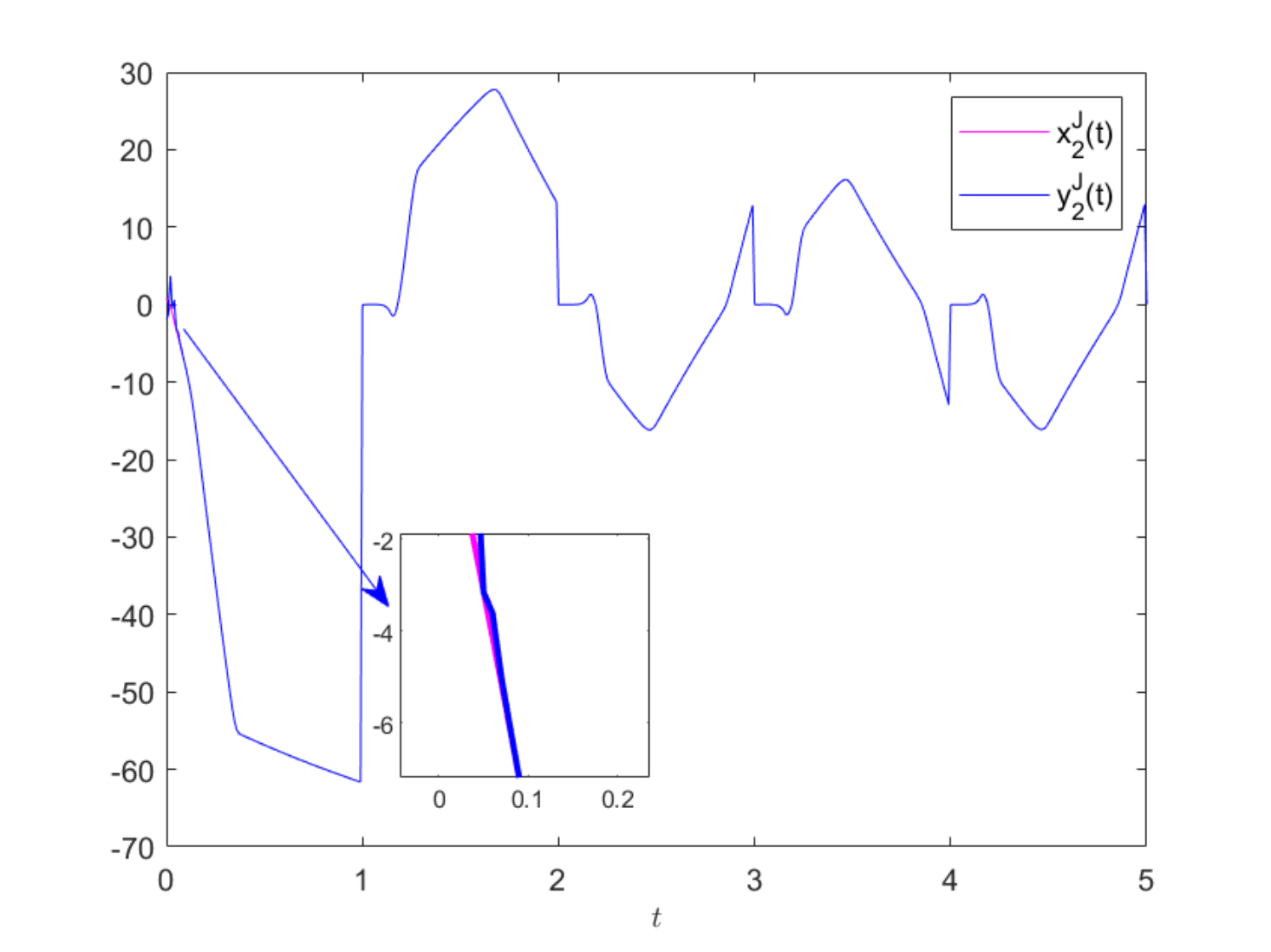}
}\noindent
\subfloat[]{
\includegraphics[width=0.49\linewidth,height=1.2in]{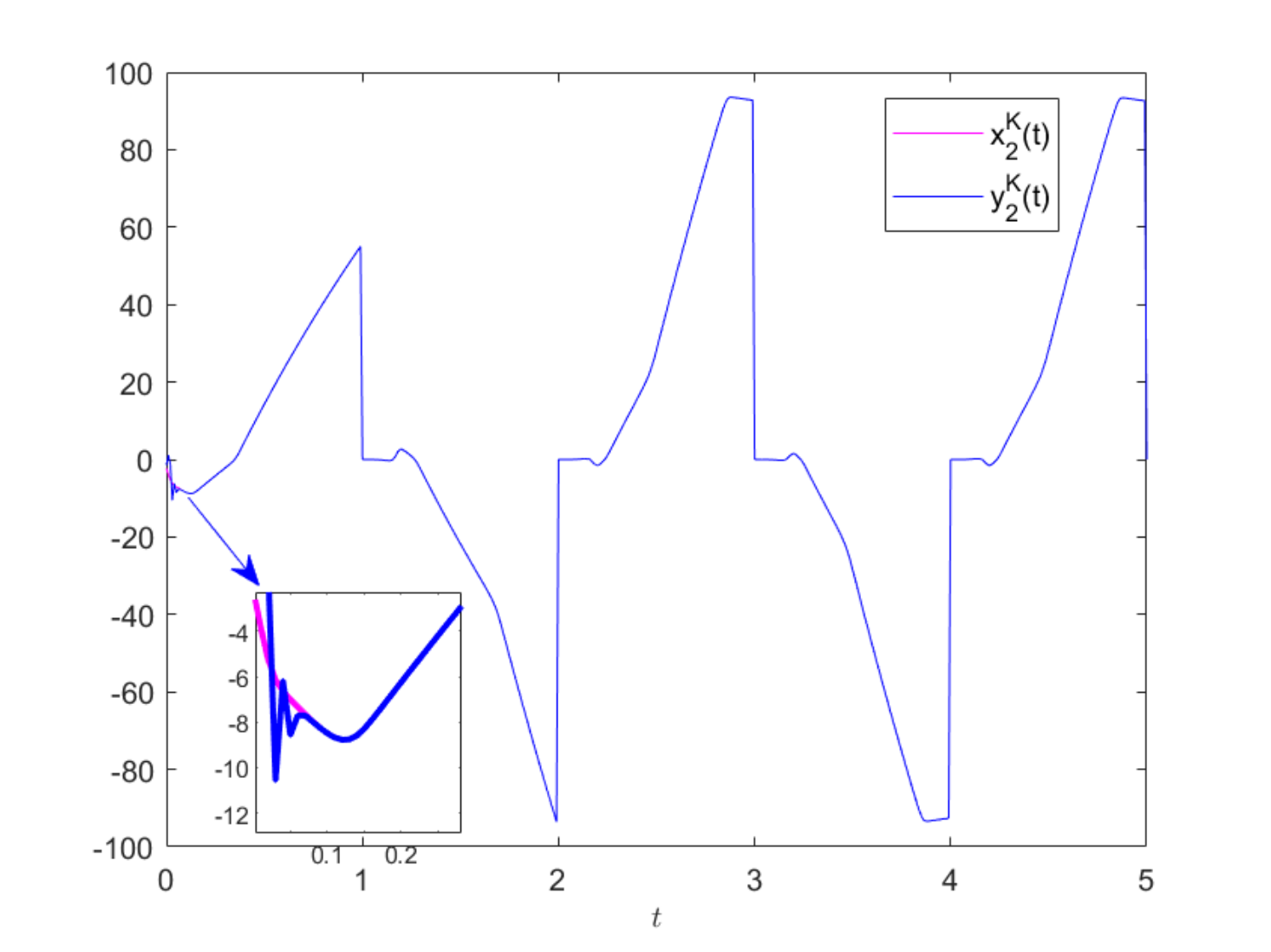}
}
\end{minipage}
\vspace{-0.12in}	
\caption{Trajectories of the real and imaginary parts of drive-response system with controllers (4.3) and (4.4) when $p=2$.}
\vspace{0.0in}		
\end{figure}

Moreover, Fig. 2 shows the whole process of FXTSNY under one-norm. It can be concluded that the error system reaches $0$ in a fixed time, which once implies that before $T=0.2$ one can achieve FXTSNY of the drive-response system. We can calculate the $T_{1}\approx1.491$ according to the settling time estimated in \eqref{3.3}, which is larger than the real synchronization time but is smaller than conventional estimation. It shows that Theorem 3.1 is correct. In addition, according to the equation \eqref{3.11} the settling time $T_{2}\approx2.628 $ can be derived from Corollary 1. By comparison, the estimated settling time derived from Lemma 2.1 is more accurate than Remark 2.1.

\vspace{3ex}

\noindent \textbf{Example 2}. The two-dimensional UCQVMNNs with mixed delays were given by the drive system \eqref{4.1} and the response system \eqref{4.2}, which consider the fixed-time synchronization of Theorem 2. 
The parameters take the same as in Example 1. 
The initial conditions of \eqref{4.1} and \eqref{4.2} are chosen as
$\bm \phi_{1}(s)=-3.8+4.8i-1.5j+1.6k,\ \bm \phi_{2}(s)=-4+0.9i-2.3j+2k,\ 
\bm \psi_{1}(s)=1.2+2.5i-1.93j+0.7k,\ \bm \psi_{2}(s)=-2.3+2.8i-1.4j-2.3k,\ s\in[-0.7,0]$.

Correspondingly, using the 2-dimensional controller by \eqref{3.13} and we have
adaptive rules for $k_{1p}, k_{2p}$, $k_{3p}(p=1,2)$ are define in \eqref{3.14} with coefficients $k_{21}=-0.26, k_{22}=-0.2,$ $k_{31}=-10.25$, $k_{32}=-9.52.$

The evolution of real and imaginary parts of system \eqref{4.1} and \eqref{4.2} are presented in Fig. 5 under the effect of controller \eqref{3.13}. And these two figures show that The drive-response system can achieve FXTSYN in a very short time under the action of the controller \eqref{3.13}. Next, Matlab drawing verification is carried out mainly for the initial error greater than 1.

\begin{figure}[H]
\centering
\vspace{-0.15in}
\begin{minipage}{1\linewidth }
\subfloat[]{
\includegraphics[width=0.49\linewidth,height=1.7in]{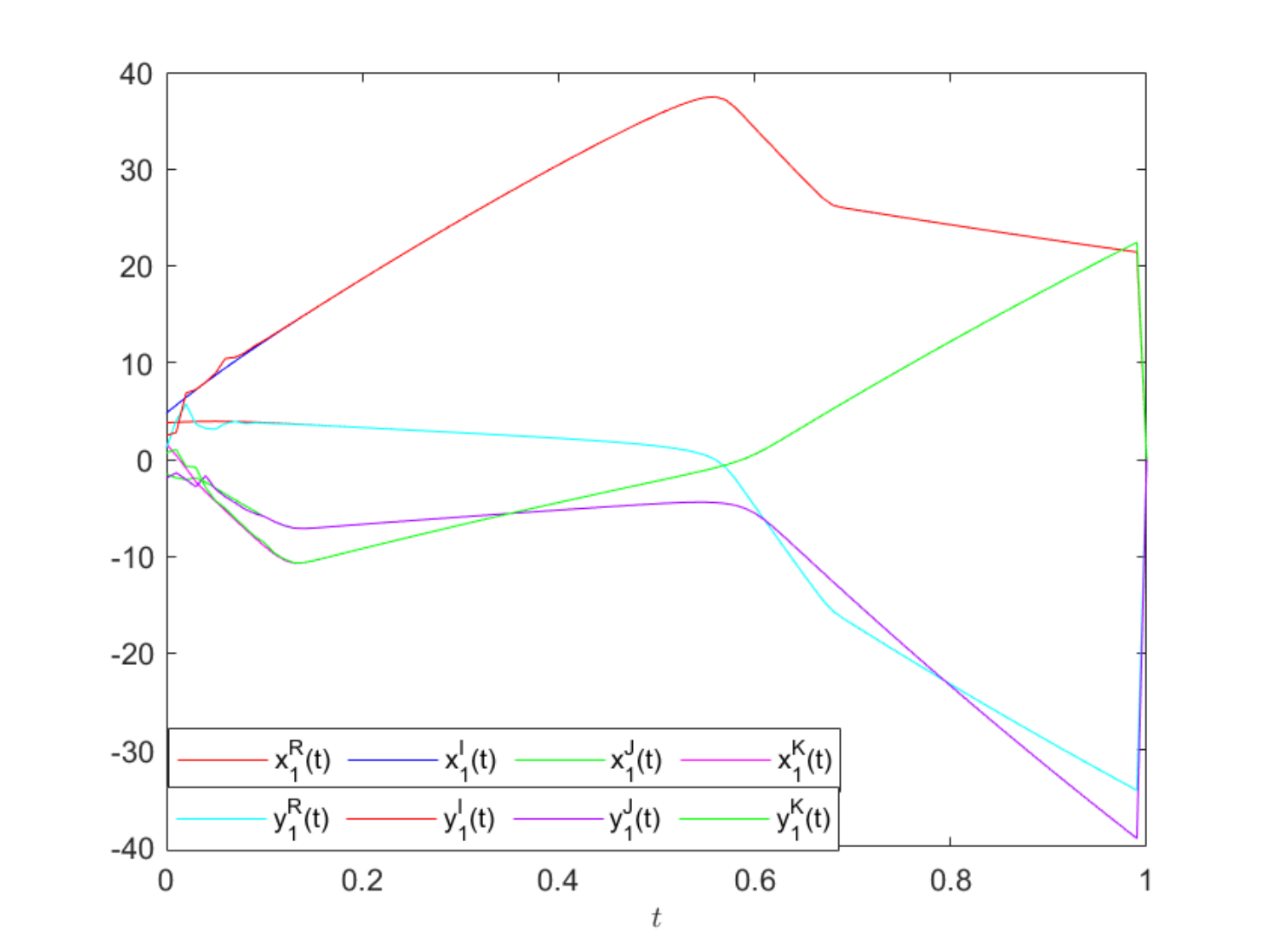}
}\noindent
\subfloat[]{
\includegraphics[width=0.49\linewidth,height=1.7in]{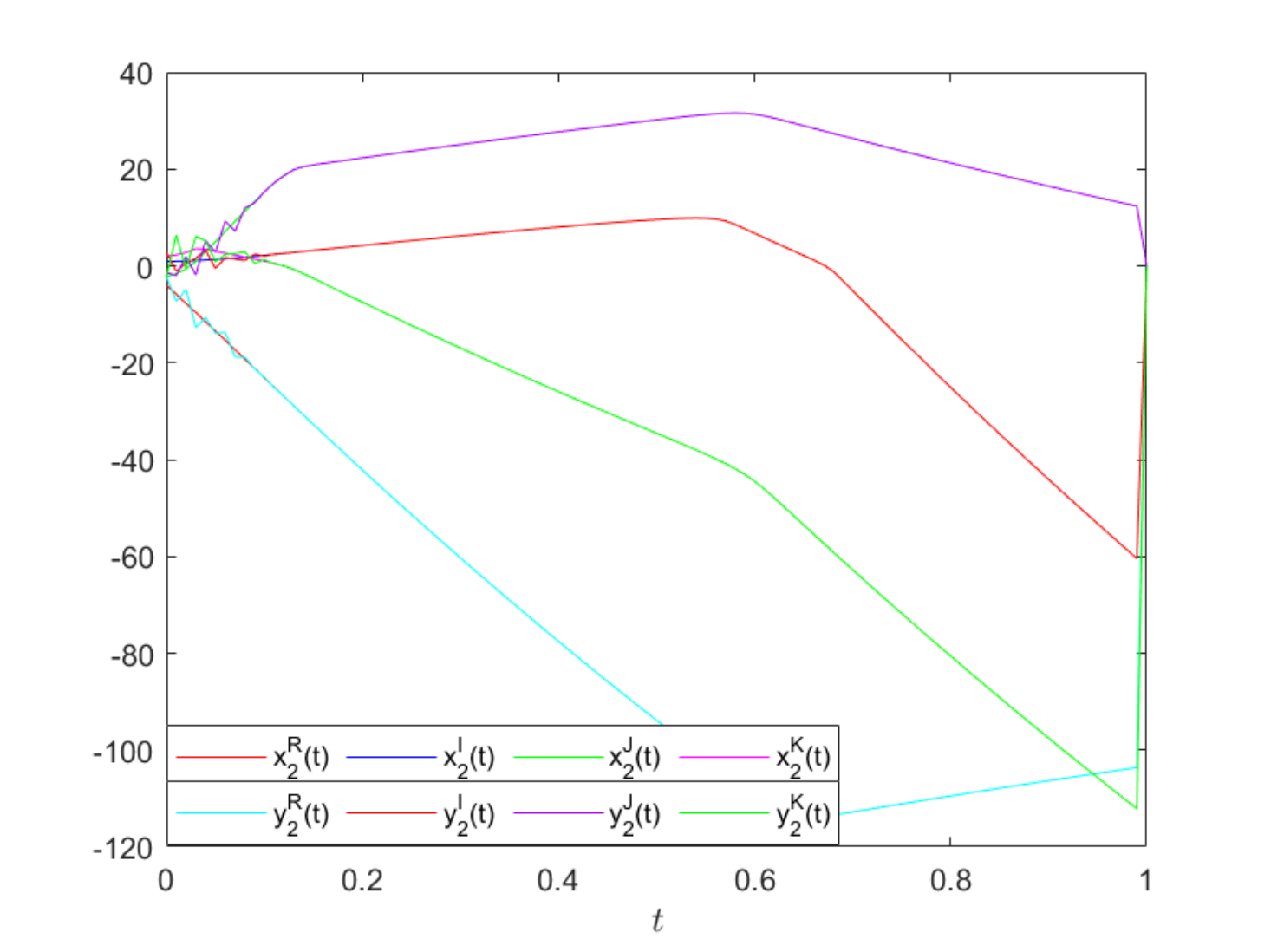}
}
\end{minipage}
\vspace{-0.10in}	
\caption{The drive-response system trajectories of $p=1,2$ with controller (3.13).}
\vspace{-0.2in}		
\end{figure}
\vspace{0.08in}

Firstly, let $\mu=40, \gamma =1.5$, if $k_{11}=37, k_{12}=130$, we can get $d=-5.1<0, \mu_{1}=5$ satisfy the condition \eqref{3.14} in Theorem 3.2. 
And the evolution trajectory of the error system with controller \eqref{3.13} is shown in Fig. 6(a), we can clearly see that the error system reaches stability in a very short time. According to Theorem 3.2, we can calculate a relatively accurate estimate of the settling time, which is $T_{4} \approx 0.160$ $(T_{3} \approx 0.065)$.

\vspace{0.1in}
\begin{figure}[H]
	\centering
	\vspace{-0.05in}
	\begin{minipage}{1\linewidth }
		\subfloat[$d<0$]{
			\includegraphics[width=0.34\linewidth,height=1.7in]{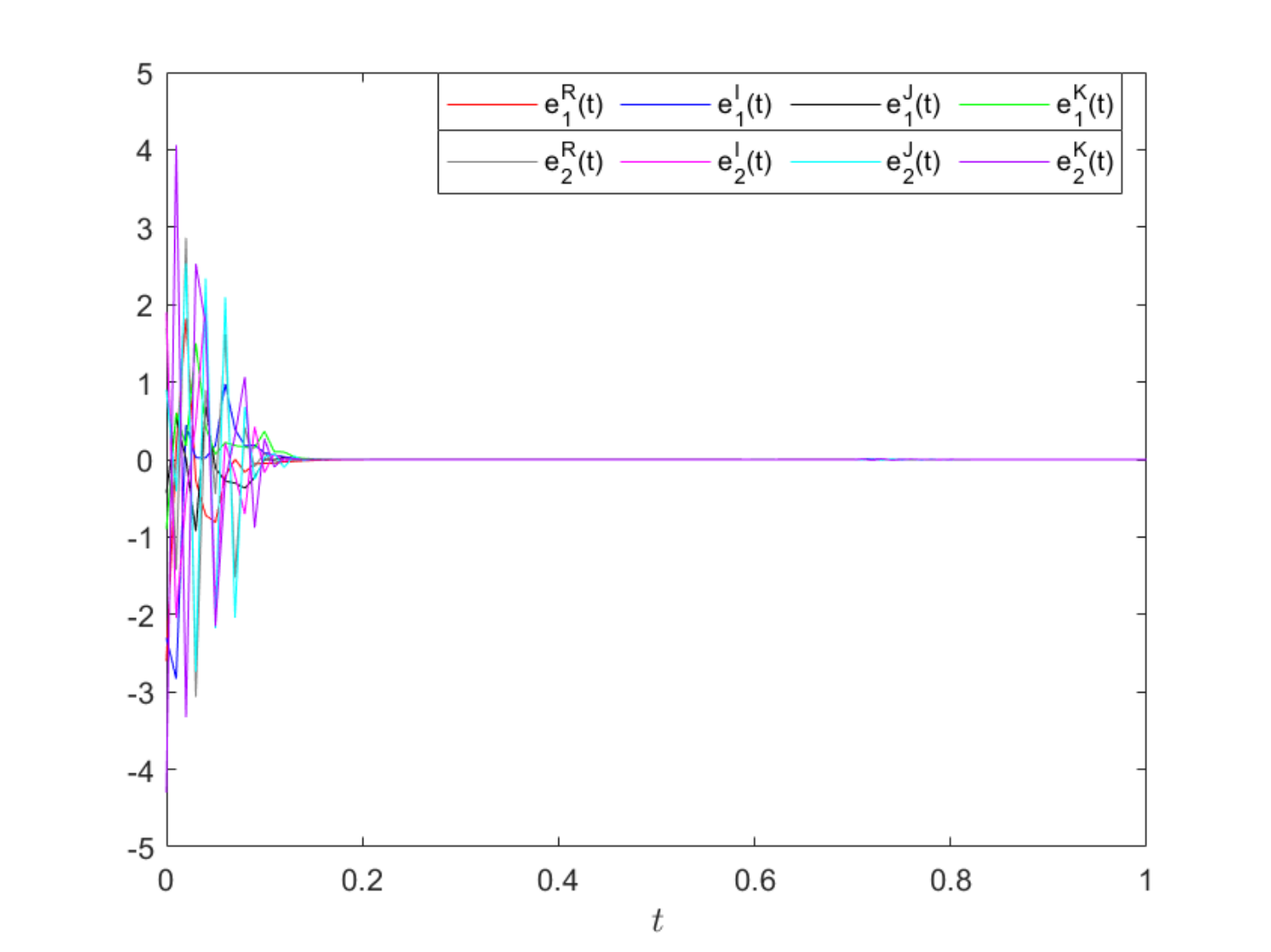}
	}\noindent
	\hspace{-0.34in}
	\subfloat[$d=0$]{
		\includegraphics[width=0.35\linewidth,height=1.7in]{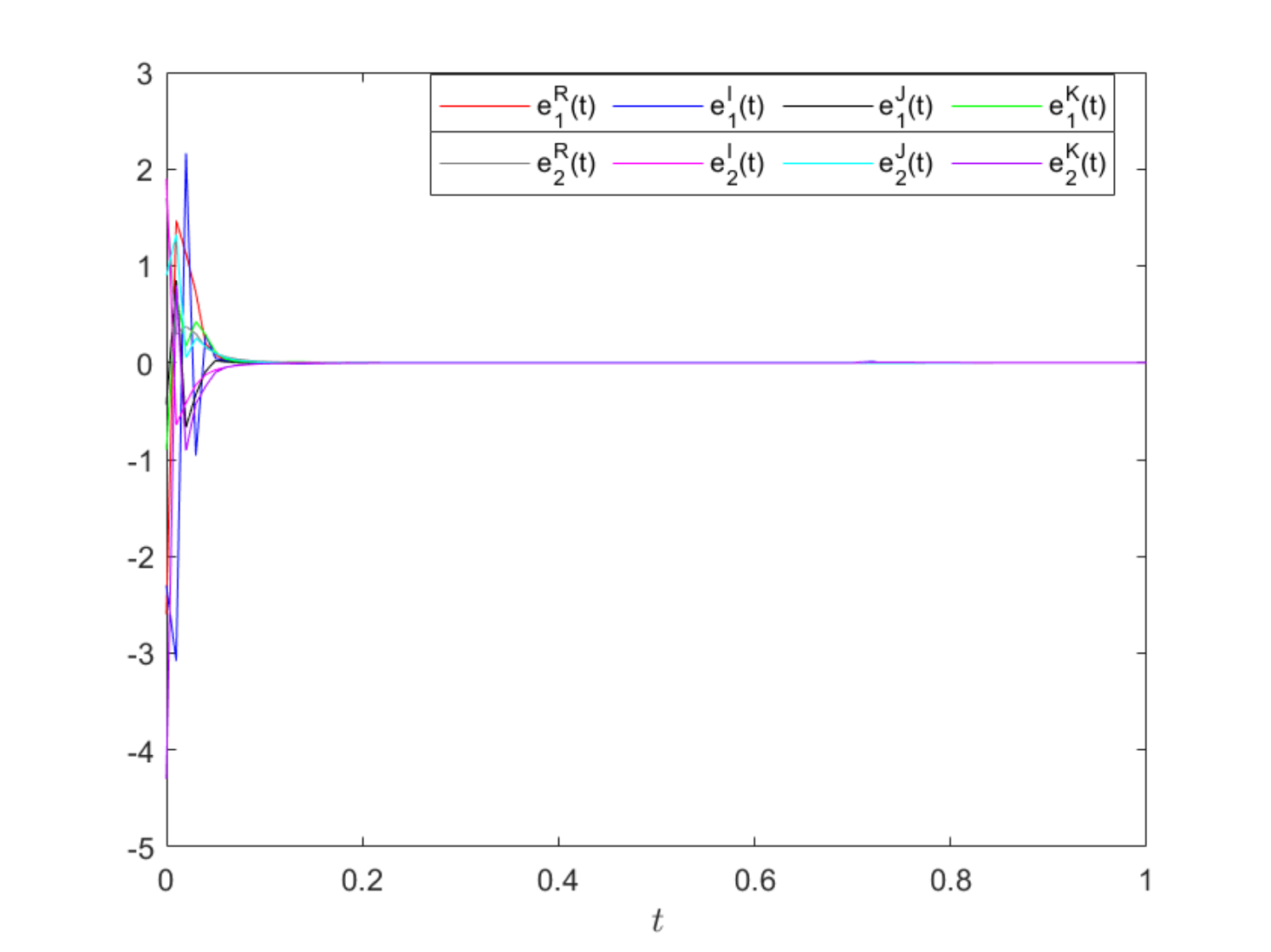}
	}
\hspace{-0.34in}
\subfloat[$d>0$]{
	\includegraphics[width=0.35\linewidth,height=1.7in]{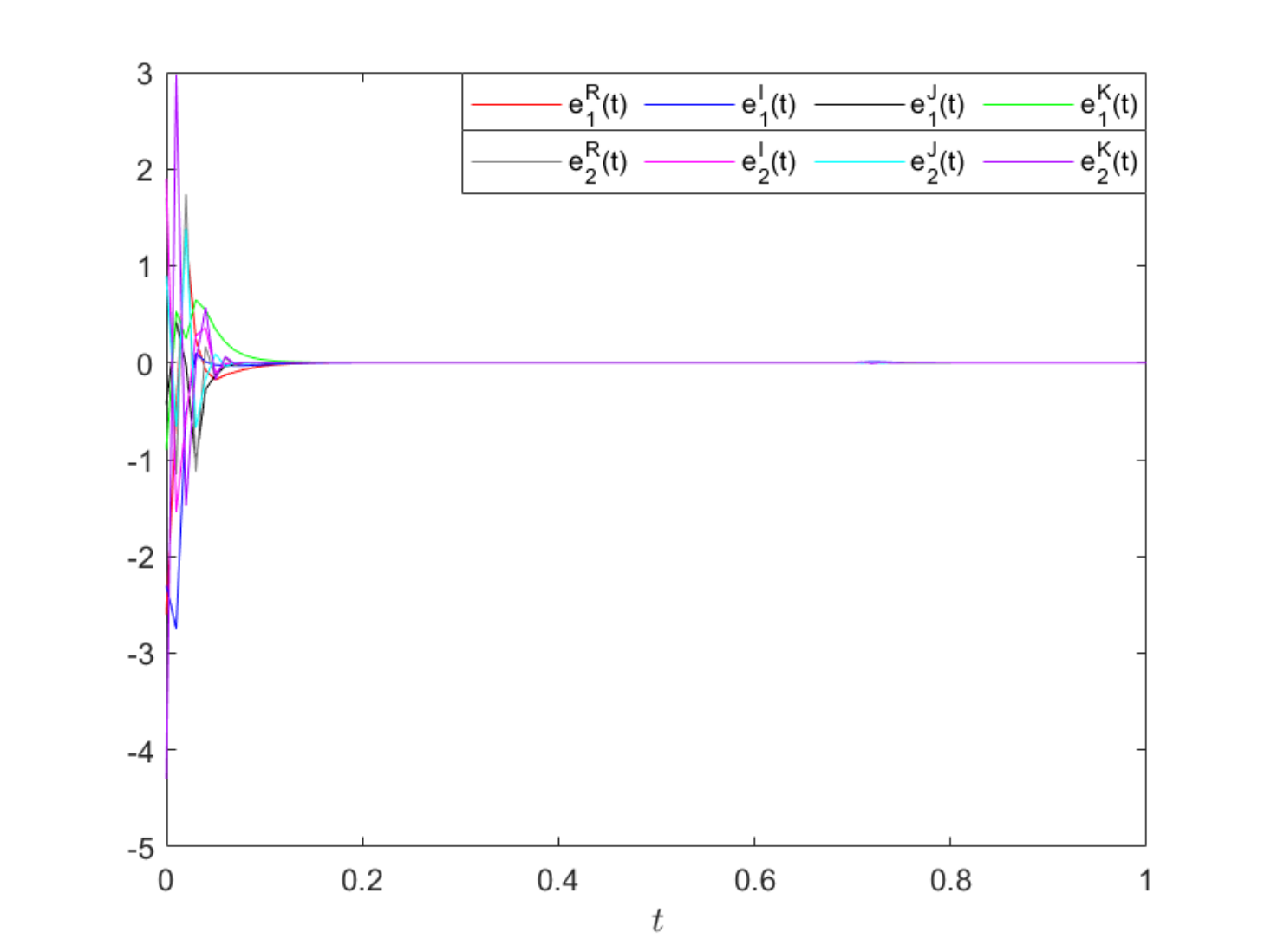}
}
\end{minipage}
\vspace{0.0in}	
\caption{Evolution of error states between networks (4.1) and (4.2) under controller (3.13). }
\vspace{-0.01in}		
\end{figure}

Secondly, let $k_{11}=50, k_{12}=23.7$, and $\mu=56, \gamma =1.5$, it's easy to can get $d=0, \mu_{1}=7$ satisfy the condition \eqref{3.14} in Theorem 3.2. Then the corresponding evolution of the error state with controller \eqref{3.13} are depicted in Fig. 6(b). Therefore, we can see from this picture that the synchronization of systems \eqref{4.1} and \eqref{4.2} can be realized within $t = 0.2$. Furthermore, $T_{4} \approx 0.131$ $(T_{3} \approx 0.048)$ according to the settling time estimated in Theorem 3.2.

Finally, if $k_{11}=33, k_{12}=130$, we can get $d=1.9>0, \mu_{1}=4$ satisfy the condition $d<\mu_{1}$. Then the dynamics of the error system with controller \eqref{3.13} are depicted in Fig. 6(c) with $\mu=32, \gamma =1.5$. Obviously, the drive system \eqref{4.1} and response system \eqref{4.2} achieve synchronization within the time $t =0.2$. And, according to the equation \eqref{3.16} in Theorem 3.2, we can calculate the settling time is $T_{4} \approx 0.382$ $(T_{3} \approx 0.087)$, which is larger than the real synchronization time, but is smaller than the conventional estimation. 

\begin{table*}[htbp] 
\centering  
\caption{Comparisons of the settling time between different $k_{11}$.}  
\scalebox{0.97}{
\begin{tabular}
{l|  c  c c c  c r}
\hline
$k_{11}$  &  31   &  33   & 34.9  &   40   &  45    &  50     \\
d     &  3.9  &  1.9  &  0    &  -5.1  & -10.1  &  -15.1  \\
$T_{3}$   & 0.071 & 0.069 & 0.067 &  0.063 &  0.059 &  0.057    \\
$T_{4}$   & 0.659 & 0.266 & 0.183 &  0.139 &  0.118 &  0.104  \\
\hline
\end{tabular}}
\end{table*}  
\vspace{0.00in}
\begin{table*}[htbp] 
\centering   
\caption{Comparisons of the settling time between different $\mu$.}  
\scalebox{0.97}{
\begin{tabular}
{l|  c  c c c  c r}
\hline
$\mu$     &  8    &  16   &   24   &   32   &  40    &   48     \\
$\mu_{1}$   &  1    &   2   &   3    &    4   &   5    &   6      \\
$T_{3}$   & 0.216 & 0.129 & 0.093 &  0.072 &  0.059 &  0.050    \\
$T_{4}$   & 0.321 & 0.216 & 0.167 &  0.137 &  0.118 &  0.103  \\
\hline
\end{tabular}}
\end{table*}  
\vspace{0.02in}
\begin{table*}[htbp] 
\centering  
\caption{Comparisons of the settling time between different $\gamma$.}  
\scalebox{0.97}{
\begin{tabular}
{l|  c  c c c  c r}
\hline
$\gamma$  &  1.3    &  1.4   &   1.5   &   1.6   &  1.7    &  1.8    \\
$T_{3}$   & 0.091  & 0.072   &  0.059  &  0.051  &  0.045  &  0.040    \\
$T_{4}$   & 0.158  & 0.134   & 0.118   &  0.106  &  0.096  &  0.089  \\
\hline
\end{tabular}}
\end{table*}  

From \eqref{3.15} and \eqref{3.16}, we know that parameters $k_{2p}, k_{3p} (p=1,2)$ have no impact on the settling time. Tables $1$-$3$ show the comparisons of the settling time for different $k_{11}$ ($k_{12}=130, \mu=40,\gamma=1.5$), $\mu$ ($k_{12}=130, k_{11}=45,\gamma=1.5$), and $\gamma$ ($k_{12}=130, k_{11}=45,\mu=40$). It is not hard to find that with controller parameters $k_{11}, \mu$ and $\gamma$ increasing, the settling time $T_{3}$ and $T_{4}$ decrease. According to these, we can choose more proper parameters as the requirements are met. In a nutshell, it is not difficult to get through the comparison of the above two examples. The settling time estimation value obtained through Theorem $3.2$ is more accurate than Theorem $3.1$.

\vspace{2ex}

\noindent \textbf{Example 3}. Consider the $128 \times 128$ pixels color image pattern "Baboon" that is depicted in Fig. 7(a). Additionally, create UCQVMNNs that have the form of the system \eqref{2.2} for associatively remembering the color image. Considering the computational complexity, we divided the image into $64$ blocks for processing (Fig. 7(b)), each block with $16\times16$ pixels. Therefore, each block needs $256$-dimensional neurons to store it. So we need to design UCQVMNNs \eqref{2.2} composed of $256$ neurons that have a $256$-dimensional equilibrium point storing the colors of the pattern.

\begin{figure}[htpb]
	\centering
	\begin{minipage}{1\linewidth }
		\centering
		\subfloat[Baboon]{
			\label{baboon}
			\includegraphics[width=0.25\linewidth,height=1.45in]{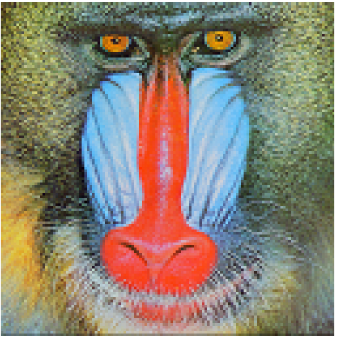}
	}\noindent
	\hspace{0.85in}
	\subfloat[ $8\times8$ blocks]{
		\includegraphics[width=0.25\linewidth,height=1.45in]{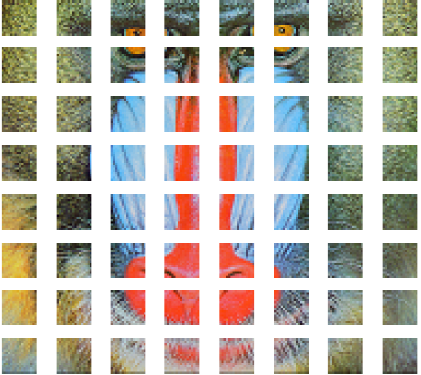}
	}
\end{minipage}
\vspace{-0.0in}
\caption{The original image and its segmented image.}
\vspace{0.0in}		
\label{}
\end{figure}

The original image (Fig. 7(a)) has an additional missing, which causes extremely sparse initial values for the response system \eqref{2.2}. The specific values are chosen as $\bm \phi_{1}(t)=(0.8863i+0.5373j+0.4902k, 0, 0.8824i+0.5059j+0.4157k,\cdots,0.8235i+0.3804j+0.3255k,0)\in \mathbb{H}^{256},$ $\bm \phi_{2}(t)=(0,\cdots,0.6549i+0.2275j+0.2029k,\cdots,0.8039i+0.3490j+0.3412k, \cdots)\in \mathbb{H}^{256},$ $\cdots$, $\bm \phi_{64}(t)=(0,\cdots,0.5021i+0.2196j+0.3294k,\cdots)\in \mathbb{H}^{256}$.  The weight coefficient are $\bm C=(\bm c_{pq})=\rm diag(10,\cdots,10)$ $\in \mathbb{H}^{256\times256}$, $d=0.01,$ $p,q=1,2,\cdots,256,$ and
\begin{align*}
&\bm a_{pq}(\cdot)=
\begin{cases}
-0.2+0.2i-0.5j+0.4k,\quad\ q< p,
\\
2+0.3i-0.2j+0.3k, \qquad \quad q= p,
\\
-0.1+0.2i+0.3j-0.5k, \quad \ \ q>p,
\end{cases}
\end{align*}
\vspace{-0.1in}
\begin{align*}
	&
	\bm b_{pq}(\cdot)=
	\begin{cases}
		0.04+0.04i-0.03j+0.05k,\quad\ q<p,
		\\
		0.04-0.05i+0.05j-0.03k, \quad \ q\geqslant p,
	\end{cases}
\end{align*}
Take the activation functions are
$\bm f_{q}(\bm x_{q}(t))=0.06(|\bm x_{q}(t)+2|-|\bm x_{q}(t)+1|),$
$\bm g_{q}(\bm x_{q}(t))=0.05(|\bm x_{q}(t)+2|-|\bm x_{q}(t)+1|),$
$\bm h_{q}(\bm x_{q}(t))=0.01(|\bm x_{q}(t)+2|-|\bm x_{q}(t)+1|),$ where $\bm x_{q}(t)\in \mathbb{H}, q=1,2,...,256$. The equilibrium points corresponding to the $64$ small blocks color image are $\bm x^{*}_{1}=(0.3608i+0.3216j+0.1490k, 0.4706i+0.4000j+0.1686k,\cdots, 0.6157i+0.6431j+0.4275k)\in \mathbb{H}^{256},$  $\cdots$, $\bm x^{*}_{64}=(0.4941i+0.5333j+0.4235k, 0.4549i+0.4549j+0.3725k, \cdots,$ $0.2471i+0.2392j+0.2235k)\in \mathbb{H}^{256}$.
Under the positive role of the controller \eqref{3.13}, the process of image restoration is the process of the system \eqref{2.2} to reach the equilibrium state, and the time used for restoration is the time used for the system to achieve equilibrium.

\begin{figure}[H]
\centering
\vspace{0.05in}
\begin{minipage}{1\linewidth }
\subfloat[Miss 80\%]{
\label{baboon}
\includegraphics[width=0.2\linewidth,height=1.2in]{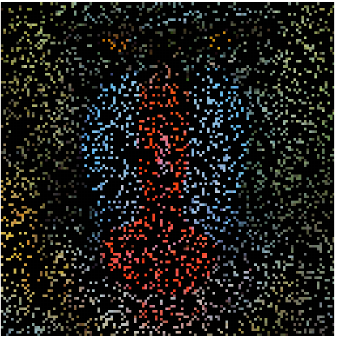}
}\noindent
\hspace{-.2in}
\subfloat[0.04s]{
\includegraphics[width=0.2\linewidth,height=1.2in]{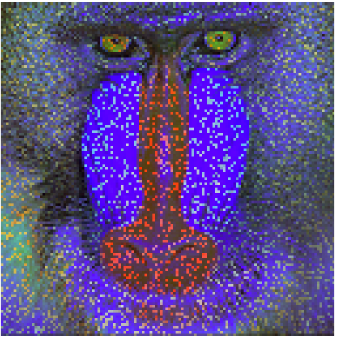}
}
\hspace{-.2in}
\subfloat[0.06s]{
\label{0.06s}
\includegraphics[width=0.2\linewidth,height=1.2in]{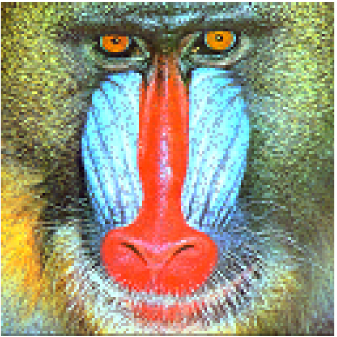}
}
\hspace{-.2in}
\subfloat[0.1s]{
\includegraphics[width=0.2\linewidth,height=1.2in]{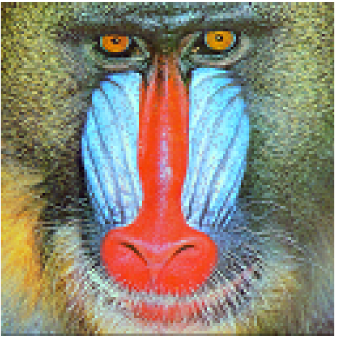}
}
\hspace{-.2in}
\subfloat[0.3s]{
	\includegraphics[width=0.2\linewidth,height=1.2in]{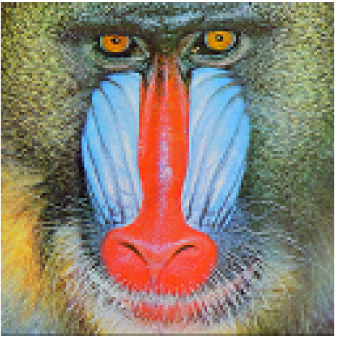}
}
\end{minipage}
\vspace{0.0in}	
\caption{Color image completion results on Image (7). (a) is the missing image where the ratio of missing pixels is $80\%$. (b) is the recovery image ( T=0.04s). (c) is the recovery image ( T=0.06s). (d) is the recovery image ( T=0.1s). (e) is the recovery image ( T=0.3s).}
\vspace{-0.02in}		
\end{figure}

\vspace{-0.05in}
\begin{figure}[H]
	\centering
	\vspace{-0.05in}
	\begin{minipage}{1\linewidth }
		\subfloat[Noise 80\%]{
			\label{baboon}
			\includegraphics[width=0.2\linewidth,height=1.2in]{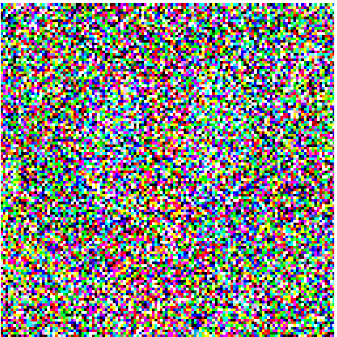}
	}\noindent
	\hspace{-0.18in}
	\subfloat[0.04s]{
		\includegraphics[width=0.2\linewidth,height=1.2in]{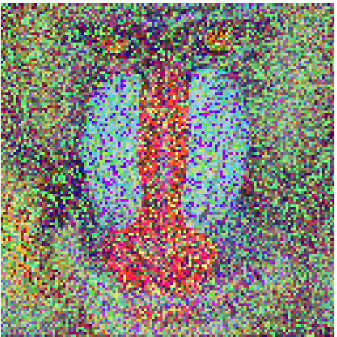}
	}
\hspace{-0.18in}
\subfloat[0.06s]{
	\label{0.06s}
	\includegraphics[width=0.2\linewidth,height=1.2in]{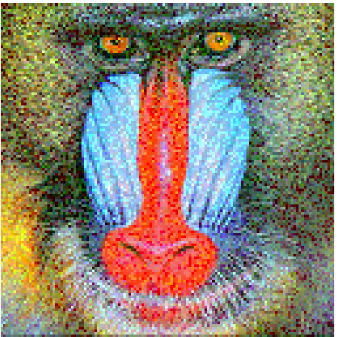}
}
\hspace{-0.18in}
\subfloat[0.1s]{
\includegraphics[width=0.2\linewidth,height=1.2in]{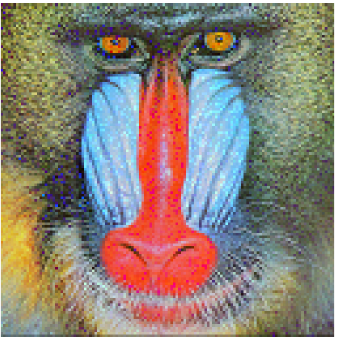}
}
\hspace{-.18in}
\subfloat[0.3s]{
\includegraphics[width=0.2\linewidth,height=1.2in]{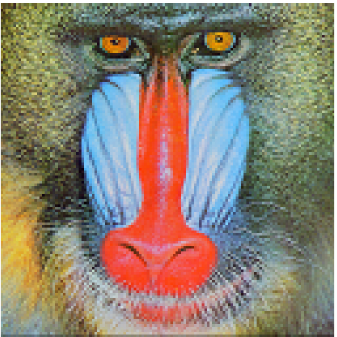}
}
\end{minipage}
\vspace{0.0in}	
\caption{Color image completion results on Image (7). (a) is the "salt and pepper" noise image where the noise density is $80\%$. (b) is the recovery image ( T=0.04s). (c) is the recovery image ( T=0.06s). (d) is the recovery image ( T=0.1s). (e) is the recovery image ( T=0.3s). }
\vspace{-0.02in}		
\end{figure}

It can be seen from Fig. 8(a), 8(b), 8(c), 8(d), and 8(e) that system \eqref{2.2} can reach the equilibrium state in a short time under the effect of the controller \eqref{3.13}. It also shows that the color images can get rapid recovery when the ratio of missing pixels is $80\%$. From Fig. 9(a), 9(b), 9(c), 9(d), and 9(e), we can see that the system can reach the equilibrium point quickly under the controller \eqref{3.13}, that is, the "Baboon" image under the premise of adding $80\%$ density "salt and pepper" noise can be quickly recovered.

Therefore, as long as the suitable controller is designed, we can use UCQVMNNs to quickly recover color images from any missing or noise state.
Thus it shows us the high efficiency of UCQVMNNs in dealing with highdimensional image restoration problems, and the controller of the theorem has a significant practical application.

\section{Conclusions}

The FXTSNY is discussed in this paper for a class of UCQVMNNs with mixed delays. Since the decomposition technology is generally accompanied by a more complex derivation process about quaternion. As a result, the quaternion-valued state is considered as a whole, with one-norm employed to achieve FXTSNY of UCQVMNNs smoothly and directly. Then based on the Lyapunov stability theorem, set-valued map, and differential inclusion theorem, we effectively deal with the system discontinuity caused by the memristor's weight coefficient in drive-response systems using the measurable selection theorem. Furthermore, sufficient conditions for the FXTSNY of delayed UCQVMNNs are proposed using the inequality technique and the Lyapunov stability theorem. In addition, different estimation settling times are obtained based on various FXTSNY criteria. It is simple to prove that Theorem 3.2's estimation value is more accurate than Theorem 3.1's. Finally, three numerical examples are provided to demonstrate the validity and effectiveness of theoretical results, as well as the practical value in high-dimensional color image processing. 

Preassigned-time synchronization is a more flexible synchronization method that is used in conjunction with the non-commutativity of the quaternion. It should be worthwhile to investigate new methods for studying the preassigned-time synchronization of QVMNNs in future work.

\section{Acknowledgement}

This work was supported by University of Macau (MYRG2022-00108-FST), Science and Technology Development Fund, Macao S.A.R (FDCT/0036/2021/AGJ).



\bibliographystyle{unsrt}
\bibliography{mybibfile}
\end{document}